\documentclass[%
 reprint,
nofootinbib,
 amsmath,amssymb,
 aps,
]{revtex4-2}

\usepackage{hyperref}
\hypersetup{
    colorlinks = true,
    linkcolor = blue,
    anchorcolor = blue,
    citecolor = blue,
    filecolor = blue,
    urlcolor = blue
    }

\usepackage{subcaption} 
\usepackage{xcolor}
\usepackage{graphicx}
\usepackage{dcolumn}
\usepackage{bm}

\usepackage{scalerel} 
\DeclareMathOperator*{\uint}{\scalerel*{\rotatebox{8}{$\!\scriptstyle\int\!$}}{\int}\hspace{-1pt}} 

\newcommand{\nnn}{\nonumber \\}
\newcommand{\rs}{{\bar r_s}}
\newcommand{\dr}{\mathrm d\bar r\,}
\newcommand{\As}{\mathcal A_s}

\newcommand{\beeq}{\begin{equation}}
\newcommand{\eneq}{\end{equation}}
\newcommand{\rbr}[1]{\left(#1\right)}
\newcommand{\sbr}[1]{\left[#1\right]}
\newcommand{\angbr}[1]{\left\langle#1\right\rangle}

\newcommand{\st}{_{\mathrm{st}}}
\newcommand{\pmz}{{_{\pm2}}}
\newcommand{\habla}{\hat\nabla}
\newcommand{\cale}{\mathcal{E}}

\begin{document}

\preprint{APS/123-QED}

\title{General Relativistic Effects \\ in Weak Lensing Angular Power Spectra}

\author{Nastassia Grimm\textsuperscript{1}} \email{ngrimm@physik.uzh.ch} \author{Jaiyul Yoo\textsuperscript{1,2}}%
 \email{jyoo@physik.uzh.ch}
 \affiliation{\textsuperscript{1}Center for Theoretical Astrophysics and Cosmology, Institute for Computational Science, University of Z\"urich, Winterthurerstrasse 190, CH-8057, Z\"urich, Switzerland \\
 \textsuperscript{2}Physics Institute, University of Z\"urich,\\ Winterthurerstrasse 190,  CH-8057, Z\"urich, Switzerland
 }



\date{\today}

\begin{abstract}
Advances in upcoming weak lensing surveys pose new challenges for an accurate modeling of the lensing observables. The wide sky coverage of Euclid makes angular scales down to $l_\mathrm{min}=10$ accessible. At such large angular scales, general relativistic effects manifest themselves, and the lensing magnification cannot be correctly described by the standard lensing convergence only. {The impact of line-of-sight velocities on the magnification angular power spectrum, referred to as the Doppler magnification, is already well recognized in literature. In particular, it was suggested that the Doppler magnification could be extracted by measurements of both cosmic shear and magnification. In this work, we point out two previously neglected aspects with respect to this method.} Firstly, the impact of the Doppler magnification is reduced through non-vanishing cross terms with the standard lensing convergence. This is particularly relevant when the sources are averaged over a bin of width $\Delta z\approx 0.1$, such as in Euclid's tomographic weak lensing survey. {Secondly, general relativistic potential terms slightly enhance the signal. We present numerical calculations of all relativistic effects in the weak lensing angular power spectra on large scales.} 
\end{abstract}

\maketitle


\section{Introduction}

Upcoming weak lensing surveys conducted by the Vera C.~Rubin Observatory (formerly LSST;~\cite{2004AAS...20510802S}), the Nancy Grace Roman Space Telescope (formerly WFIRST;~\cite{green2012widefield}) and the Euclid satellite~\cite{Laureijs:2011gra} will provide us with an exciting opportunity to learn more about cosmology and fundamental physics. In addition to a drastic improvement in measurement precision compared to the current generation, these surveys will lead to an impressive increase in the covered survey-area: for instance, Euclid will in total cover 15,000 square degrees on the sky, ten times more than KiDS~\cite{Kuijken:2015vca} which is one of the most important weak lensing surveys conducted so far. However, these advances in measurements might also lead to potentially biased conclusions if not all observational and theoretical challenges are adequately addressed. From an observational point of view, a number of well-known systematics needs to be properly accounted for (see e.g.~\cite{Mandelbaum:2017jpr} for a review). At the same time, these advances also require a more precise theoretical modeling of weak lensing observables, as  approximations that have previously been sufficient need to be reconsidered.

With its large survey area, Euclid will be able to measure previously inaccessible large scales with $l_\mathrm{min}=10$~\cite{Blanchard:2019oqi}, thereby probing a regime where additional general relativistic effects manifest themselves.  As pointed out in~\cite{Yoo:2018qba}, the standard formalism for the calculation of weak lensing observables suffers from gauge-dependencies, indicating that it does not accurately account for all physical effects. To resolve this issue, fully gauge-invariant weak lensing formalisms have been presented in~\cite{Yoo:2018qba,Grimm:2018nto,Schmidt:2012ne}. In particular, lensing magnification effects are not properly described by the standard lensing convergence alone. The correct description is given by the distortion in the angular diameter distance which, apart from angular distortions leading to the standard lensing convergence, also accounts for distortions in the radial direction and the observed redshift on the light cone. 

Indeed, the impact of peculiar line-of-sight velocities on the observed radial coordinate and redshift, referred to as the \textit{Doppler magnification} or \textit{Doppler lensing}, is well known in literature. 
It was first derived in~\cite{Bonvin:2008ni}, and further work studied the implications for cosmological observables~\cite{Amendola:2016saw, Andrianomena:2018aad, Duniya:2016gcf, Montanari:2015rga, Bacon:2014uja, Bonvin:2016dze, Bolejko:2012uj}. As concluded in~\cite{Bacon:2014uja}, the Doppler magnification dominates over the standard convergence for small redshifts $z\lesssim 0.3$ and hence can be measured directly through the magnification of sizes (see~\cite{Schmidt:2011qj, Casaponsa:2012tq, Alsing:2014fya}). For larger redshifts, however, the standard convergence is more significant. In~\cite{Bonvin:2016dze}, it was proposed to measure the Doppler magnification via the dipole in the cross-correlation of galaxy sizes and number counts. Here, they concluded that it dominates up to $z \approx 0.5$, but also noted that for larger redshifts the standard convergence is not negligible and these effects thus need to be modeled together.

To disentangle the standard convergence and Doppler magnification, another method was {noted in~\cite{Bacon:2014uja} and} proposed in~\cite{Amendola:2016saw} with respect to the Euclid survey. While the magnification has a contribution from the line-of-sight velocities, cosmic shear is to linear order related to the standard convergence only. Hence, measuring the magnification in addition to the cosmic shear angular power spectrum would provide a novel way of measuring peculiar velocities, making it an important complementary probe. More specifically,~\cite{Amendola:2016saw} concluded that the Doppler magnification would be measurable up to a redshift of $z=0.6$ if Euclid reaches a precision of 10\%, and up to $z=1$ if a precision of 1\% can be achieved. 

{However, several issues complicate the measurement.} 
First of all, the width of the bin smears out the Doppler magnification. Secondly, non-vanishing cross terms between the Doppler magnification and the standard lensing convergence lead to a further reduction of the signal. {While these cross terms have been previously neglected~\cite{Bonvin:2008ni, Amendola:2016saw, Bacon:2014uja}, we show in Sec.~\ref{Section:Results} that they lead to a large reduction of the velocity signal for bins corresponding to Euclid's tomographic weak lensing survey. We also demonstrate, by applying a unique source redshift at $z=0.6$, that this reduction still remains significant even for the idealized case of an infinitely thin redshift bin.} 

General relativistic corrections to the standard lensing convergence consist not only of the Doppler magnification, but also of additional potential terms evaluated at the source position and along the line of sight. 
The existence of these additional \textit{GR potential terms} is recognized in literature~\cite{Yoo:2018qba, Grimm:2018nto, Schmidt:2012ne, Duniya:2016gcf, Bacon:2014uja, Yoo:2016vne, Bonvin:2005ps, Bonvin:2016dze}. In principle, along with the velocity terms they lead to general relativistic corrections in the angular power spectrum of the lensing magnification and its cross power spectrum with cosmic shear. However, their contribution was assumed to be negligible in~\cite{Amendola:2016saw, Bacon:2014uja}. 
{Here, additionally to investigating the impact of Doppler-convergence cross terms, we also quantify the magnitude of GR potential terms for Euclid's tomographic weak lensing survey.} In such a setting with redshift bins of non-negligible width, GR potential terms are relevant compared to the Doppler magnification{, although the combined signal is indeed small compared to cosmic variance.} 
 
This paper is structured as follows: in Sec.~\ref{Section:Preliminaries}, we revise the necessary preliminaries for our work, including the fully relativistic expressions for the weak lensing observables (Sec.~\ref{Section:WLO}) and expressions for scalar perturbations and their growth functions (Sec.~\ref{Section:Perturbations}).  In Sec.~\ref{Section:FullSky}, we present the analytical results for the fully relativistic weak lensing angular power spectra{ and explain the relation between them.} 
Then, in Sec.~\ref{Section:Results}, we present our numerical evaluations for the magnification angular power spectrum (Sec.~\ref{ResultsD}) and the cross angular power spectrum between the magnification and shear E-modes (Sec.~\ref{ResultsE}). 
We summarize and conclude on our results in Sec.~\ref{Section:Conclusion}. In Appendix~\ref{Appendix:VectorCalc}, we present some basic vector calculus identities in spherical coordinates that are used throughout this work, and in Appendix~\ref{Appendix:kdirection} and~\ref{Appendixgammakn}, we present details for the analytical calculations of weak lensing angular power spectra that are referred to where appropriate.

\section{Preliminaries} \label{Section:Preliminaries}

In Sec.~\ref{Section:WLO}, we review the theoretical expressions for the weak lensing observables and their general relativistic corrections. In particular, we use the linear-order expressions derived in~\cite{Yoo:2018qba, Grimm:2018nto}, which are gauge-invariant and include all general-relativistic effects (see also~\cite{Schmidt:2012ne}). 
Together with the relations for perturbation variables and their power spectra in Sec.~\ref{Section:Perturbations}, they will serve as a basis for the analytical and numerical investigation of the fully relativistic weak lensing angular power spectra presented in the subsequent sections. 

\subsection{Weak lensing observables in general relativity} \label{Section:WLO}
Weak gravitational lensing effects are commonly described via the distortion of the angular source positions on the sky (see e.g.~\cite{Bartelmann:1999yn}). In this simple treatment, which we refer to as the \textit{standard formalism}, the shape and size distortion of an infinitesimal image observed at a two-dimensional angle $\boldsymbol n=(\theta,\phi)$ and redshift $z_s$ is described by the Jacobian matrix:
\beeq
\mathcal A_{ij}
\equiv\delta_{ij}-\frac{\partial\alpha_i}{\partial n_j}
\equiv\begin{pmatrix}1&0\\0&1\end{pmatrix}-\begin{pmatrix}
{\kappa}\st+{\gamma_1}\st&{\gamma_2}\st\\ {\gamma_2}\st&\kappa\st-{\gamma_1}\st\end{pmatrix}\,, \label{defampmatrix}
\eneq
where $\boldsymbol\alpha$ is the two-dimensional deflection angle. 
The Jacobian matrix $\mathcal A_{ij}$ is usually referred to as amplification or distortion matrix. The trace $\kappa\st$ is called the convergence and describes the magnification of images, while ${\gamma_1}\st$ and ${\gamma_2}\st$ are the shear components and describe the shape distortions.\footnote{Note that all these weak lensing observables depend on the observed line-of-sight direction $\mathbf n$, and redshift $z_s$, e.g.~$\kappa\st\equiv\kappa\st(\mathbf n,z_s)$. This is typically omitted in our notation throughout this paper, although the dependence on $\mathbf n$ is sometimes written explicitly as a way of emphasizing it.} The deflection angle $\boldsymbol\alpha$ is given by the gradient of the projected lensing potential $\psi$,
\beeq
\alpha_i=\frac{\partial\psi}{\partial n_i}\,,\qquad\psi=\int_0^{\rs }\mathrm d\bar r\rbr{\frac{\rs-\bar r}{\rs \bar r}}2\Psi(\bar r)\,, \label{mostsimpleform}
\eneq
where $\Psi(\bar r)$ is the Newtonian potential and $\rs$ is the comoving distance to the source associated to the observed source redshift $z_s$,
\beeq
\bar r_s=\int_0^{z_s}\frac{\mathrm dz}{H(z)}\,. \label{rs}
\eneq

However, as discussed in detail in~\cite{Yoo:2018qba}, the standard formalism faces several problems and in particular leads to gauge dependencies. 
In essence, these gauge dependencies arise from the fact that the standard formalism relies on a description of light propagation in global FLRW coordinates. Observables described in global coordinates would require a global observer looking at the universe as a whole and are thus not physically meaningful. To correctly describe weak lensing quantities, observables need to be described in the observer rest frame and compared to the intrinsic size and shape in the source rest frame. Such a gauge-invariant description of all weak lensing observables including all general relativistic effects was given in~\cite{Yoo:2018qba, Schmidt:2012ne, Grimm:2018nto}. Furthermore, relativistic corrections to the magnification were also described in~\cite{Bonvin:2008ni, Bonvin:2005ps, Yoo:2016vne}.

For the fully general linear-order results for the weak lensing observables, i.e.~including scalar, vector and tensor modes and without choosing a certain gauge, we refer the reader to these papers. Here, we state the results adopting the Newtonian gauge and ignoring vector and tensor perturbations. We consider the perturbed FLRW metric
\begin{align}
\mathrm ds^2=-a^2(\eta)(1+2\Psi)\mathrm d\eta^2+a^2(\eta)(1-2\Psi)\delta_{\alpha\beta}\mathrm dx^\alpha\mathrm dx^\beta\,,
\end{align}
where $\eta$ is the conformal time and $a(\eta)$ is the expansion scale factor. Additionally, we consider the fact that the observer's motion is perturbed. While the timelike ($u^\mu u_\mu=-1$) four-velocity of any comoving observer is given in an unperturbed FLRW universe by $\bar u^\mu=(1/a,0)$, its perturbed value is given by
\beeq
u^\mu=\frac 1a \rbr{1-\Psi,V^\alpha}\,,
\eneq
where $V^\alpha$ is the observer's peculiar velocity. To quantify magnification effects, the standard convergence $\kappa\st$ is replaced by the distortion in the angular diameter distance,
\beeq
\delta D=-\kappa_{\mathrm{st}}+\delta z+\frac{\delta r}{\bar r_z}-\Psi_s \,. \label{deltaD} 
\eneq
Here, the \textit{standard convergence} that is generalized to account for the observer's motion (see~\cite{Yoo:2018qba}) is given by
\begin{align}
 \kappa_\mathrm{st}=&-{V_{\parallel}}_o+\frac{n_\alpha\delta x^\alpha_o}{\rs}+\int_0^{\rs}\mathrm d\bar{r}\rbr{\frac{\bar{r}_s-\bar{r}}{\bar{r}_s\bar r}}\habla^2\Psi\,, \label{kappast}
\end{align} 
where $n^\alpha$ is the line-of-sight direction, $\habla_\alpha$ is the angular gradient for which we give explicit expressions in Appendix~\ref{Appendix:VectorCalc}, and $\delta x^\alpha_o$ is the spatial coordinate lapse at the observer position. The additional contributions to $\delta D$ arise from the distortion of the redshift,
\begin{align}
\delta z=\left(\mathcal H\delta\eta\right)_o+\left({V}_\parallel-\Psi\right)_o^s-2\int_0^{\bar r_s}\mathrm d\bar r\,\Psi'\,, \label{deltaz}
\end{align}
the distortion of the radial coordinate of the source,
\begin{align}
\delta r=n_\alpha\delta x^\alpha_o+\delta\eta_o-\frac{\delta z}{\mathcal H_s}+2\int_0^{\bar r_s}\mathrm d\bar r\, \Psi \,, \label{deltar}
\end{align}
and an additional perturbation $-\Psi_s$ at the source position arising from correctly relating global coordinates to the rest frame of the source galaxy. Again, we want to emphasize that $\kappa\st$ is neither observable nor gauge-invariant. Only the full quantity $\delta D$ correctly quantifies measurable magnification effects. Hence, we will refer to the distortion in the angular diameter distance $\delta D$ as the \textit{magnification} hereafter.\footnote{The lensing magnification $\delta D$ is degenerate with the intrinsic source brightness if the latter is unknown. This degeneracy in lensing is unrelated to gauge-invariance. A measurement of $\delta D$ is possible given sufficient knowledge of the intrinsic distribution of sizes or magnitudes~\cite{Schmidt:2011qj}, while a measurement of the gauge-dependent quantity $\kappa_\mathrm{st}$ is never possible.}

Summing up all contributions, we can rewrite Eq.~\eqref{deltaD} into
\begin{align}
\delta D=&\mathcal A_s\Psi_o+\left(\mathcal H_o\mathcal A_s+\frac{1}{\bar r_s}\right)\delta\eta_o+\frac{V_{\parallel o}}{\bar r_s\mathcal H_s} \nnn
&-\int_0^{\bar{r}_s}\mathrm d\bar{r}\,\left(\frac{\bar{r}_s-\bar{r}}{\bar{r}_s\bar r}\right)\habla^2\Psi +\mathcal A_s V_{\parallel s}-(\mathcal A_s+1)\Psi_s  \nnn
&-2\mathcal A_s\int_0^{\bar r_s}\mathrm d\bar r\,\Psi'+\frac{2}{\bar r_s}\int_0^{\bar r_s}\dr\Psi \,, \label{deltaDscalar}
\end{align}
{where we defined the dimensionless quantity $\mathcal A_s\equiv 1-(\mathcal H_s\bar r_s)^{-1}$ for ease of notation.} The first line of this expression consists of observer terms that ensure the gauge-invariance of the expression, but will only appear in the monopole and the dipole of the magnification angular power spectrum. The first term in the second line corresponds to the standard convergence, while the line-of-sight velocity term is a relativistic correction referred to as the \textit{Doppler magnification} (see e.g.~\cite{Andrianomena:2018aad, Bacon:2014uja}). The remaining terms of this expression are additional potential terms evaluated at the source position and along the line of sight which we call the \textit{GR potential terms}.\footnote{With the expression \textit{GR potential terms}, we will refer only to the additional terms without the standard convergence, even though it is technically a general relativistic potential term itself.} 

Note that the contribution of the spatial coordinate lapse $\delta x^\alpha_o$ 
cancels out in the observable $\delta D$. However, the contribution of the time coordinate lapse $\delta \eta_o$ given by
\beeq
\delta\eta_o=-v_o\,,\qquad v^{,\alpha}\equiv- V^\alpha\,,
\eneq
 where $v$ is the velocity potential, is not vanishing. In literature, both the spatial and the temporal lapses $\delta x^\alpha_o$ and $\delta\eta_o$ are typically set to zero. However, this would correspond to a specific gauge-choice -- the comoving-synchronous gauge -- which is incompatible with e.g.~the Newtonian gauge chosen in this work, where all gauge degrees of freedom are already fixed. Hence, while the contributions of $\delta x^\alpha_o$ from $\kappa\st$ and $\delta r$ cancel out, the non-vanishing contribution of $\delta\eta_o$ needs to be considered, as omitting observer terms can break the gauge-invariance and further lead to unphysical artifacts such as infrared divergences~\cite{Biern:2016kys}. 
 
The expressions for the shear components in the fully general relativistic description are given by
\begin{align}
{\gamma_i}=\Phi^{\alpha\beta}_i \int_0^{\bar{r}_s}\mathrm d\bar{r}\rbr{\frac{\bar{r}_s-\bar{r}}{\bar r_s\bar{r}}}\habla_\alpha\habla_\beta\Psi \,,
\label{gamma}
\end{align}
where $\Phi_i^{\alpha\beta}$ is defined as
\beeq
\Phi_1^{\alpha\beta}={\theta^\alpha\theta^\beta-\phi^\alpha\phi^\beta}\,,\qquad  \Phi_2^{\alpha\beta}={\theta^\alpha\phi^\beta+\phi^\alpha\theta^\beta}\,. \label{defPhii}
\eneq
Here, $\theta^\alpha$ and $\phi^\alpha$ are two vectors orthogonal to the line-of-sight direction $n^\alpha$, with their explicit expressions given in Appendix~\ref{Appendix:VectorCalc}. Hereafter the subscript $i$, appearing in quantities related to the shear components, will always refer to $i=1,2$. 
Unlike the magnification, the shear components take the same expressions as in the standard formalism for the scalar modes in the Newtonian gauge. Relativistic effects arise only for higher-order calculations~\cite{Bernardeau:2011tc}, or when taking vector and tensor modes into account~\cite{Schmidt:2012ne, Yoo:2018qba, Grimm:2018nto, Schmidt:2012nw}.

\subsection{Perturbation variables and their power spectra} \label{Section:Perturbations}
The effect of weak lensing on observed images is determined by the perturbations of the FLRW metric and the peculiar motion. To numerically compute the power spectra of weak lensing observables, we thus need precise knowledge of the power spectra of these perturbation variables. 
For our work we apply the power spectrum $P_{m,o}(k)$, describing the matter inhomogeneities $\delta_o(\mathbf{k})$ at $a_o=1$, produced by \texttt{CLASS} \cite{Blas:2011rf} for a flat $\Lambda$CDM universe. Given today's matter power spectrum $P_{m,o}(k)$, its past evolution in a $\Lambda$CDM universe can be described by a scale-independent growth function $D(a)$,
\beeq
\delta(k,a)=D(a)\delta_o(k)\,,\qquad P_m(k,a)=D^2(a) P_{m,o}(k)\,,
\eneq 
given by the ordinary hypergeometric function ${_2}F_1$  (see e.g.~\cite{Shirasaki}),
\beeq
D(a)=\frac{\tilde D(a)}{\tilde D(1)}\,,\quad \tilde D(a)=a\, {{_2}F_1\sbr{\frac 13,1,\frac{11}{6},\frac{a^3(\Omega_m-1)}{\Omega_m}}}\,,
\eneq
where we assumed $w=-1$ for the dark energy equation of state and normalized the growth function to unity at $a_o=1$ today.
To compute the weak lensing angular power spectra, we further need to know the growth functions of the velocity potential and the scalar potential $\Psi$. This is, in essence, achieved by applying the ADM equations (see e.g.~\cite{Arnowitt:1962hi, Noh:2004bc}) in the comoving gauge to obtain the curvature power spectrum $P_\zeta(k)$ and the growth function of the metric perturbations in that gauge, and then apply the transformation to the Newtonian gauge used in this work. We refer to~\cite{Biern:2016kys, Yoo:2016tcz} for the details, and only state the results here.

The time-independent curvature perturbation $\zeta(\mathbf x)\equiv \zeta(\mathbf x,\eta)$ is related to the density perturbation $\delta_o(\mathbf x)$ today as
\beeq
\zeta(\mathbf x)=C\Delta^{-1}\delta_o(\mathbf x)\,,\qquad C=-\mathcal H^2 f\Sigma D\,,
\eneq
where $C$ is a constant, and we defined
\beeq
f\equiv\frac{\mathrm d\ln D}{\mathrm d\ln a}\,,\qquad \Sigma\equiv 1+\frac{3}{2}\frac{\Omega_m}{f}\,.
\eneq
The scalar potential $\Psi$ and line-of-sight velocity $V_\parallel=V_\alpha n^\alpha$ are related to the curvature perturbation $\zeta$ as
\beeq
\Psi(\mathbf x, \eta)=D_\Psi(\eta)\zeta(\mathbf x)\,,\qquad V_\parallel(\mathbf x, \eta)=D_V(\eta )\partial_\parallel\zeta(\mathbf x)\,, \label{scalarpert}
\eneq
where the solutions for $D_\Psi(\eta)$ and $D_V(\eta)$ are given by
\beeq
D_V=\frac{1}{\mathcal H\Sigma}\,,\qquad D_\Psi=\mathcal HD_V-{1}=-\frac{1}{2}(D_V'+1)\,. \label{GrowthFunctions}
\eneq

\section{Analytical expressions for the fully relativistic angular power spectra} \label{Section:FullSky}

Upcoming surveys such as Euclid will cover a large survey area, and thus make large angular scales available. Therefore, a precise modeling of weak lensing observables over the whole celestial sphere is necessary, which has been studied in various literature (see e.g.~\cite{Stebbins:1996wx, Duniya:2016gcf, Schmidt:2012ne, Kitching:2016zkn, Hu:2000ee}). In particular,~\cite{Schmidt:2012ne} and~\cite{Duniya:2016gcf} have considered general relativistic effects, including potential terms, in their full-sky calculation of the magnification angular power spectrum. Here, we revise the theoretical formalism based on (spin-weighted) spherical harmonics, and state the results for the magnification angular power spectrum (Sec.~\ref{SecDeltaDfull}) and the shear E-mode angular power spectrum as well as the magnification E-mode cross angular power spectrum (Sec.~\ref{SecEBfull}).

A signal $A(\mathbf n)$ observed on the whole sky can be decomposed using the spherical harmonics decomposition,
\beeq
A(\mathbf{ n})=\sum a^A_{lm}Y_{lm}(\mathbf{ n})\,,\qquad a^{A}_{lm}=\int\mathrm d\Omega\, A(\mathbf{ n})Y^\ast_{lm}(\mathbf{ n})\,,
\eneq
where we choose the convention
\begin{align}
&Y_{lm}(\theta,\phi)=\sqrt{\frac{(2l+1)}{4\pi}\frac{(l-m)!}{(l+m)!}}P_l^m(\cos\theta)e^{im\phi}\,,\nnn &P_l^m(x)=(-1)^m(1-x^2)^{m/2}\frac{\mathrm d^m}{\mathrm dx^m}P_l(x)\,, \label{defYlm}
\end{align}
for the spherical harmonics $Y_{lm}(\theta,\phi)$ and the associated Legendre polynomials $P_l^m(x)$. The angular power spectrum $C^{A}(l)$ is related to the spherical harmonics coefficients $a^A_{lm}$ as
\beeq
\angbr{a^A_{lm}a^{A\ast}_{l'm'}}=\delta_{ll'}\delta_{mm'}C^{A}(l)\,.
\eneq
We will work with the quantities $A(\mathbf n,\mathbf k)$ and $a^A_{lm}(\mathbf k)$ denoting the contribution of a single Fourier mode,
\begin{align}
&A(\mathbf n)\equiv\int\frac{\mathrm d^3\mathbf k}{(2\pi)^3}A(\mathbf k,\mathbf n)\,,\quad a^A_{lm}=\int\frac{\mathrm d^3\mathbf k}{(2\pi)^3}a^A_{lm}(\mathbf k)\,,\nnn 
&a^{A}_{lm}(\mathbf k)\equiv\int\mathrm d\Omega\, A(\mathbf k,\mathbf{ n})Y^\ast_{lm}(\mathbf{n})\,. \label{alm}
\end{align}
Throughout this section, the coordinate $\bar r$ in expressions such as $\Psi(\mathbf k,\bar r)$ will be used to denote the conformal time coordinate $\eta(\bar r)=\eta_o-\bar r$.
 
\subsection{Fully relativistic angular power spectrum of the magnification} \label{SecDeltaDfull}

To obtain the angular power spectrum of the magnification given in Eq.~\eqref{deltaDscalar}, we first compute the contribution from a single Fourier mode. Splitting $\delta D(\mathbf k, \mathbf n)$  into observer terms ($o$) and non-observer contributions from the standard convergence ($\kappa$), the Doppler magnification ($v$) and GR potential terms ($p$),
\beeq
\delta D(\mathbf k,\mathbf n)\equiv\big(\delta D^{o}+\delta D^\kappa+\delta D^{v}+\delta D^{p}\big)(\mathbf k,\mathbf n)\,,
\eneq
we obtain
\begin{align}
&\delta D^o(\mathbf k,\mathbf{ n})=\As\Psi(\mathbf k,0)+\left(\mathcal H_o\As+\frac{1}{\bar r_s}\right)\delta\eta(\mathbf k,0) \nnn
&\phantom{\delta D^o(\mathbf k,\mathbf{ n})=}+\frac{V_{\parallel}(\mathbf k,\mathbf n,0)}{\bar r_s\mathcal H_s}\,, \nnn
&\delta D^\kappa(\mathbf k,\mathbf{ n})=-\int_0^{\bar{r}_s}\mathrm d\bar{r}\,\rbr{\frac{\bar{r}_s-\bar{r}}{\bar{r}_s\bar r}}\habla^2\left(\Psi(\mathbf k,\bar r)e^{ix\mu}\right)\,, \nnn
&\delta D^{v}(\mathbf k,\mathbf{ n})=\As V_\parallel(\mathbf k,\mathbf n,\bar r_s)e^{i x_s\mu}\,, \nnn
&\delta D^{p}(\mathbf k,\mathbf{ n})=-(\As + 1)\Psi(\mathbf k,\bar r_s)e^{i x_s\mu} \nnn 
&+\frac{2}{\bar r_s}\int_0^{\bar r_s}\mathrm d\bar r\,\Psi(\mathbf k,\bar r)e^{ix\mu}-2\As\int_0^{\bar r_s}\mathrm d\bar r\,\Psi'(\mathbf k,\bar r)e^{ix\mu}\,, \label{deltaDdiffcont}
\end{align}
where we chose $\mathbf k$ to be aligned with the $z$-axis, $\mathbf k=k\mathbf e_z$, and defined $x=k\bar r$ and $\mu=\cos\theta=\mathbf n\cdot\mathbf{\hat k}$. 
Now, note that from the definition of spherical harmonics in Eq.~\eqref{defYlm} and the plane wave expansion along with the orthogonality condition the Legendre polynomials, it follows that
\begin{align}
\int\mathrm d\Omega\,Y^\ast_{lm}e^{ix\mu}=\sqrt{4\pi(2l+1)} i^l j_l(x)\delta_{m0}\,. \label{fundamentaleq}
\end{align}
We use this equation to obtain the spherical harmonic coefficient
\begin{align}
{a^{\delta D}_{lm}}(\mathbf k)=&i^l\sqrt{4\pi(2l+1)}\zeta(\mathbf k)\delta_{m0}\,\mathcal S_l^{\delta D}(k)\,, \label{adDlm}
\end{align}
 where $\mathcal S^{\delta D}_l(k)$ is the sum of:
 \begin{align}
&\mathcal S^{o}_l(k)=\As D_\Psi(0)\delta_{l0}+\left(\mathcal H_o\As+\frac{1}{\bar r_s}\right)D_V(0)\delta_{l0} \nnn
&\phantom{\mathcal S^{o}_l(k)=}+\frac{1}{3}\frac{D_V(0)k}{\bar r_s\mathcal H_s}\delta_{l1}\,, \nnn
&\mathcal S^\kappa_l(k)=l(l+1)\int_0^{\bar{r}_s}\mathrm d\bar{r}\left(\frac{\bar{r}_s-\bar{r}}{\bar{r}_s\bar r}\right)D_\Psi(\bar r)j_l(x)\,, \nnn
&\mathcal S^{v}_l(k)=\As D_V(\bar r_s)kj'_l(x_s)\,, \nnn
&\mathcal S^{p}_l(k)=
-(\As+1)D_\Psi(\bar r_s)j_l(x_s)+\frac{2}{\bar r_s}\int_0^{\bar r_s}\mathrm d\bar r\,D_\Psi(\bar r)j_l(x) \nnn
&\phantom{\mathcal S^{\delta D}_l(k)=} -2\As\int_0^{\bar r_s}\mathrm d\bar r\,D'_\Psi(\bar r)j_l(x)\,. 
\end{align}
Note that the factor $\delta_{m0}$ in Eq.~\eqref{adDlm} is a result of choosing $\mathbf k$ to be aligned with the $z$-axis. Furthermore, note that the term $\propto j_l'(x_z)$ occurs since multiplications with $i\mu$ can be transformed into partial derivatives $\partial_x$. To deal with the term  $\propto V_\parallel(0,\mathbf k,\mathbf n)$, we used the relation
\beeq
\int\mathrm d\Omega\,\mu Y^\ast_{lm}=2\sqrt{\frac{\pi}{3}}\int\mathrm d\Omega\,Y_{10}Y^\ast_{lm}=2\sqrt{\frac{\pi}{3}}\delta_{l1}\delta_{m0}\,.
\eneq
We also applied the fundamental property $\widehat\nabla^2Y_{lm}=-l(l+1)Y_{lm}$ of spherical harmonics.

Finally, we compute the scalar contribution $C^{\delta D}(l)$ to the magnification angular power spectrum,
\begin{align}
C^{\delta D}(l)&=\iint\frac{\mathrm d^3\mathbf k}{(2\pi)^3}\frac{\mathrm d^3\mathbf k'}{(2\pi)^3}\angbr{{a^{\delta D}_{lm}}(\mathbf k){a^{\delta D\ast}_{lm}}(\mathbf k')}\,.
\end{align}
Hence, $C^{\delta D}(l)$ is given by an integral over all $\mathbf k,\,\mathbf k'\,\in\,\mathbb{R}^3$, while we have only computed ${a_{lm}^{\delta D}}(\mathbf k)$ for a wave vector aligned with the $z$-axis. However, as explained in appendix~\ref{Appendix:kdirection}, a summation over all $m$ allows us to replace the general $\mathbf k$ with $k\mathbf{e_z}$, which leads to the result
\beeq
C^{\delta D}(l)=\frac{2}{\pi}\int \mathrm dk\,k^2P_{\zeta}(k)\mathcal S_l^{\delta D}(k)^2\,. \label{CdeltaD}
\eneq

As we have not assumed any approximation to derive this expression, it is valid on all scales at the linear order. Our result is mostly consistent with the result of~\cite{Duniya:2016gcf}\footnote{Note that their quantity $\mu^{-1}$ is given by $-2\delta D$, and that they have chosen an opposite sign convention for $V_\parallel$. Their opposite sign for the integral or $\propto \Psi'$ remains unexplained, although this term is small compared to the other general relativistic corrections.}, although we additionally considered the terms evaluated at the observer position. Indeed, observer terms do not contribute to any multipoles apart from the monopole $l=0$ (potential terms) and the dipole $l=1$ (line-of-sight velocity). Nevertheless, they are important to obtain a non-divergent monopole (cf.~\cite{Biern:2016kys}, where the importance of observer terms has been discussed in the context of infrared divergences in the luminosity distance).

Note that the velocity contributes to the total magnification angular power spectrum via two different terms: a velocity-velocity term $\propto \mathcal S_l^{v}(k)^2$ and a velocity-convergence cross term $\propto \mathcal S_l^{\kappa}(k)\mathcal S_l^{v}(k)$. In previous work, the cross term was considered to be negligible~\cite{Bonvin:2008ni, Amendola:2016saw}. In particular,~\cite{Bonvin:2008ni} has provided the explanation that the Doppler magnification collects Fourier modes along the line of sight, and the standard convergence those perpendicular to it. However, the product $\mathcal S_l^{\kappa}(k)\mathcal S_l^{v}(k)$ is in fact non-vanishing. While the argument involving Fourier modes perpendicular and along the line of sight is valid in a flat sky approximation, such a separation cannot be made on large angular scales, as no unique line of sight can be defined. Indeed, in our numerical result Sec.~\ref{Section:Results} we will show that for the bins of Euclid's weak lensing survey the velocity-convergence cross term significantly reduces the overall velocity effect in the magnification angular power spectrum. 

The GR potential terms contribute to the magnification angular power spectrum via the cross-term $\propto\mathcal S_l^{\kappa}(k)\mathcal S_l^{p}(k)$. The pure GR potential term $\propto\mathcal S_l^{p}(k)^2$ as well as cross-correlations between the Doppler magnification and the GR potential terms, $\propto\mathcal S_l^{v}(k)\mathcal S_l^{p}(k)$, are completely negligible.

\subsection{Shear E-modes and their cross angular power spectrum with the magnification} \label{SecEBfull}

Unlike the magnification, the shear components depend on the choice of an arbitrary basis $(\boldsymbol \theta,\boldsymbol \phi)$ orthonormal to $\mathbf n$ on the sky. For example, rotating the basis by $45^\circ$ transforms $\gamma_1$ into $\gamma_2$ and vice versa. This behavior is mathematically better described by replacing the shear components $\gamma_1$ and $\gamma_2$ with the spin-2 quantities ${_{\pm2}}\gamma\equiv \gamma_1\pm i\gamma_2$. They are given  by
\beeq
{_{\pm2}}\gamma\equiv m^\alpha_\mp m^\beta_\mp \int_0^{\bar{r}_s}\mathrm d\bar{r}\,\frac{2(\bar{r}_s-\bar{r})}{\bar{r}_s\bar r}\habla_\alpha\habla_\beta\Psi\,, \label{Eq:gammapm}
\eneq
where $m^\alpha_\pm$ describes a spin-1 basis on the sky,
\beeq
m^\alpha_\pm \equiv\frac{1}{\sqrt{2}} (\theta^\alpha\mp i\phi^\alpha)\,. 
\eneq
The spin-$2$ quantities $\pmz\gamma(\mathbf k, \mathbf n)$ are decomposed as
\beeq
a^{\gamma\pm}_{lm}(\mathbf k)\equiv\int\mathrm d\Omega\, \pmz \gamma(\mathbf k,\mathbf n)\pmz Y^\ast_{lm}(\mathbf n)\,. \label{gammadecomp} 
\eneq
For all $l\geq2$, this yields the expression
\begin{align}
a^{\gamma\pm}_{lm}(\mathbf k)=&\delta_{m0}\sqrt{\frac{4\pi(2l+1)(l+2)!}{(l-2)!}}i^l \nnn
&\times\int_0^{\bar r_s}\mathrm d\bar r \rbr{\frac{\bar{r}_s-\bar{r}}{\bar r_s\bar{r}}}\Psi(\mathbf k,\bar r)j_l(x)\,, \label{almgamma}
\end{align}
 where details of the calculation are stated in Appendix~\ref{Appendixgammakn}, along with the expressions for the spin-weighted spherical harmonics and the spin-raising and lowering operators. The quantities $a^{\gamma+}_{lm}(\mathbf k)$ and $a^{\gamma-}_{lm}(\mathbf k)$ are identical, which implies that the shear components result only in an E-mode,
\beeq
a^\cale_{lm}(\mathbf k)\equiv \frac 12\rbr{a^{\gamma+}_{lm}(\mathbf k)+a^{\gamma-}_{lm}(\mathbf k)}=a^{\gamma\pm}_{lm}(\mathbf k)\,, \label{aElm}
\eneq
while the shear B-mode is vanishing,
\beeq
a^\mathcal{B}_{lm}(\mathbf k)=\frac{1}{2i}\rbr{a^{\gamma+}_{lm}(\mathbf k)-a^{\gamma-}_{lm}(\mathbf k)}=0\,.
\eneq
A non-vanishing B-mode, distinguishable from the E-mode via its behavior under parity transformation, cannot be caused by scalar mode perturbations at linear order. Including tensor modes arising from primordial gravitational waves into the weak lensing formalism leads to a non-vanishing B-mode, which is however far too low to be measured~\cite{Schmidt:2012nw}.

Having obtained the expression for $a^\cale_{lm}(\mathbf k)$, we can rederive the well-known expression (see e.g.~\cite{Hu:2000ee}) for the shear E-mode angular power spectrum,
\begin{align}
C^\cale(l)=&\frac{1}{2l+1}\sum_{m=-l}^l\iint\frac{\mathrm d^3\mathbf k}{(2\pi)^3}\frac{\mathrm d^3\mathbf k'}{(2\pi)^3}\left\langle a^{\cale}_{lm}(\mathbf k)a^{\cale\ast}_{lm}(\mathbf k')\right\rangle \nnn
=&\frac{2}{\pi}\frac{(l+2)!}{(l-2)!}\int\mathrm dk\,k^2 P_{\zeta}(k)\rbr{ \mathcal S^{\cale}_l(k)}^2\,, \label{resultCgammal}
\end{align}
where
\begin{align}
\mathcal S_l^{\cale}(k)=\delta_{l\ge 2}\int_0^{\bar r_s}\mathrm d\bar r \left(\frac{\bar{r}_s-\bar{r}}{\bar r_s\bar{r}}\right) D_\Psi(\bar r)j_l(x) \,. \label{Slgamma} 
\end{align}
Furthermore, together with the expression for $a^{\delta D}_{lm}(\mathbf k)$ given in Eq.~\eqref{adDlm}, we immediately obtain the expression for the cross power spectrum $C^{\delta D\mathcal E}(l)$,
\begin{align}
C^{\delta D \cale}(l)=&\frac{1}{2l+1}\sum_{m=-l}^l\iint\frac{\mathrm d^3\mathbf k}{(2\pi)^3}\frac{\mathrm d^3\mathbf k'}{(2\pi)^3}\left\langle a^{\delta D}_{lm}(\mathbf k)a^{\mathcal E\ast}_{lm}(\mathbf k')\right\rangle \nonumber \\ 
=&\frac{2}{\pi}\sqrt{\frac{(l+2)!}{(l-2)!}}\int\mathrm dk\,k^2 P_{\zeta}(k) \mathcal S^{\delta D\mathcal E}_l(k)\,,
\end{align}
where 
\beeq
\mathcal S^{\delta D\cale}_l(k)\equiv\mathcal S_l^{\delta D}(k) S_l^{\cale}(k)\,.
\eneq
While the shear E-mode angular power spectrum $C^\cale(l)$ is, to linear order, unaffected by general relativistic effects, the cross angular power spectrum $C^{\delta D\cale}(l)$ is affected by the line-of-sight velocity and the additional GR potential terms in the magnification. In particular, for the same reason as discussed at the end of Sec.~\ref{SecDeltaDfull}, the contribution of the line-of-sight velocity is non-vanishing. We discuss the implications in Sec.~\ref{ResultsE}.

Note that when taking into account only the standard convergence term determined by $\mathcal S_l^\kappa(k)^2$, we recover the relation~\cite{Hu:2000ee}
\beeq
C^\cale(l)=\frac{(l+2)(l-1)}{(l+1)l}C^\kappa(l)=-\sqrt{\frac{(l+2)(l-1)}{(l+1)l}} C^{\kappa\cale}(l)\,. \label{relation}
\eneq
In~\cite{Amendola:2016saw}, it was suggested to measure the magnification in addition to the shear E-mode angular power spectrum, and then use the above relation to extract the contribution of the Doppler magnification $\propto S_l^v(k)^2$. However, as we point out in the next section, the velocity-convergence cross term significantly reduces the signal arising from the Doppler magnification for the redshift bins of Euclid's weak lensing survey. Furthermore, corrections to the magnification angular power spectrum as well as the magnification-shear cross angular power spectrum arise not only from the Doppler magnification, but also from the additional GR potential terms. 

{The pure Doppler lensing term $\propto S_l^v(k)^2$ can, in theory, be isolated by considering the cross spectrum $C^{\delta D\cale}(k)$ in addition to the auto spectra $C^{\delta D}(l)$ and $C^\cale (l)$. The cosmic shear angular power spectrum $C^\cale(l)$ is determined by the standard term $\propto S^\kappa_l(k)^2$ only, while the cross spectrum $C^{\delta D\cale}(l)$ is additionally affected by the cross terms $\propto S_l^\kappa(k)(S_l^v(k)+S_l^p(k))$ which could thus be isolated. Then, the terms $\propto S_l^\kappa(k)^2$ and $\propto 2 S_l^\kappa(k)(S_l^v(k)+S_l^p(k))$ in the magnification angular power spectrum $C^{\delta D}(l)$ would already be determined, providing a way of measuring the only remaining non-negligible term $\propto S_l^v(k)^2$. However, the impact of the terms $\propto S_l^\kappa(k)(S_l^v(k)+S_l^p(k))$ on the cross spectrum $C^{\delta D\cale}(l)$ is reduced by a factor of 2 compared to their impact on the auto spectrum $C^{\delta D}(l)$, leading to an overall small signal as illustrated in the next section. This complicates the possibility of an individual measurement.} 

Finally, we want to point out that in order to extract the impact of general relativistic contributions using Eq.~\eqref{relation}, the same redshift distribution needs to be applied for both magnification and shear measurements. Therefore, in the next section we apply the distribution and binning of Euclid's tomographic weak lensing survey to compute both the angular power spectrum of the magnification and its cross angular power spectrum with shear E-modes.

\section{Numerical Results} \label{Section:Results}

\begin{figure*} 
\begin{center}

\begin{subfigure}[b]{1\textwidth}
\hspace{1.0cm}\includegraphics[width=0.7\linewidth]{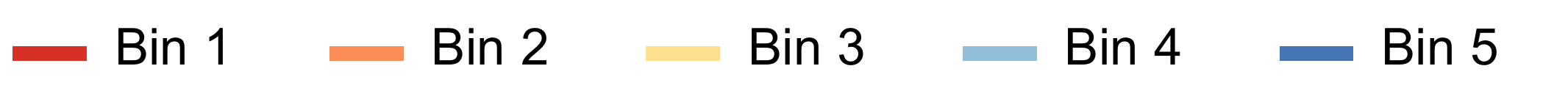} 
\end{subfigure}

\begin{subfigure}[b]{0.49\textwidth}
\centering\includegraphics[height=6.2cm]{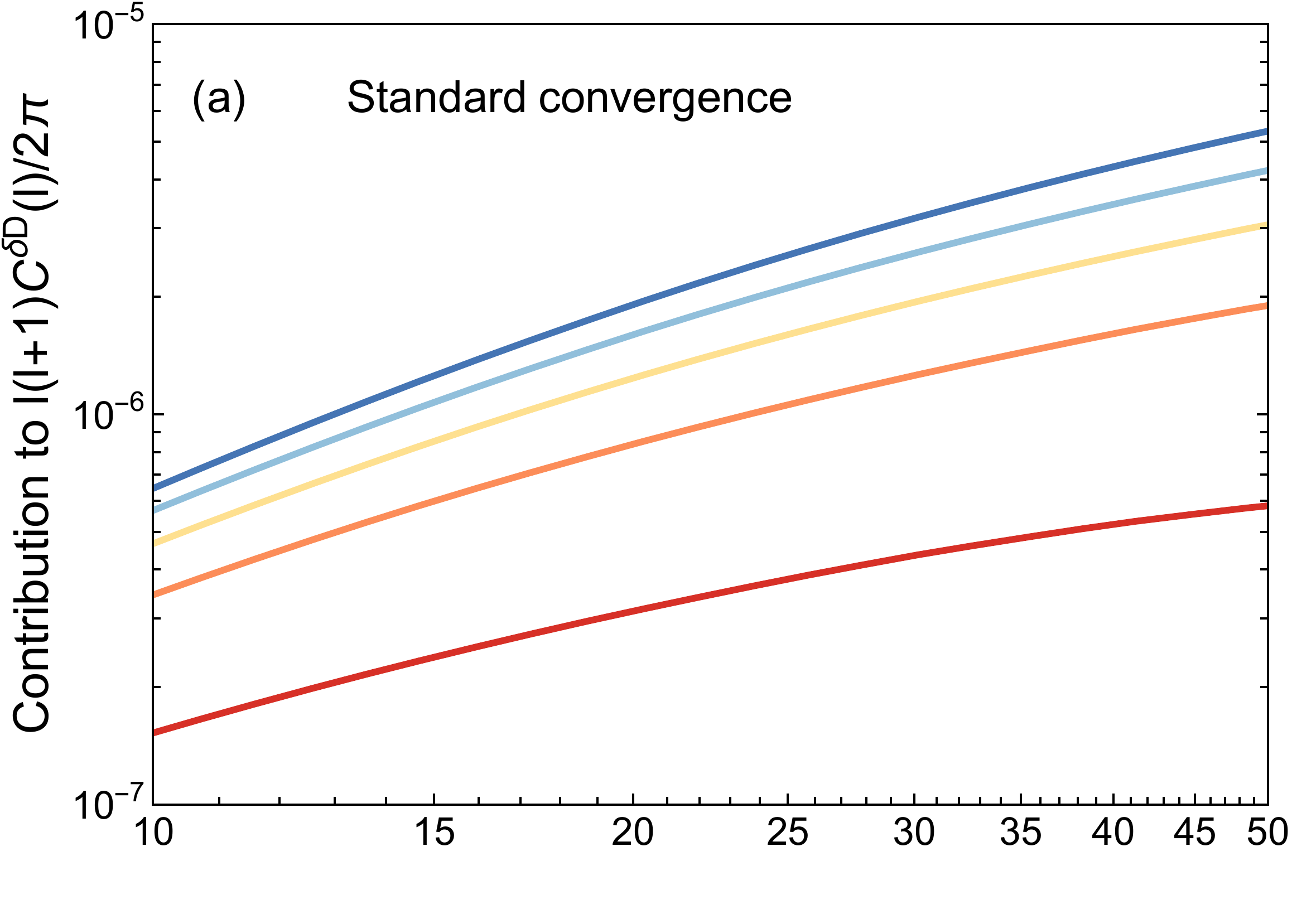}
\end{subfigure}
\begin{subfigure}[b]{0.49\textwidth}
\centering\includegraphics[height=6.2cm]{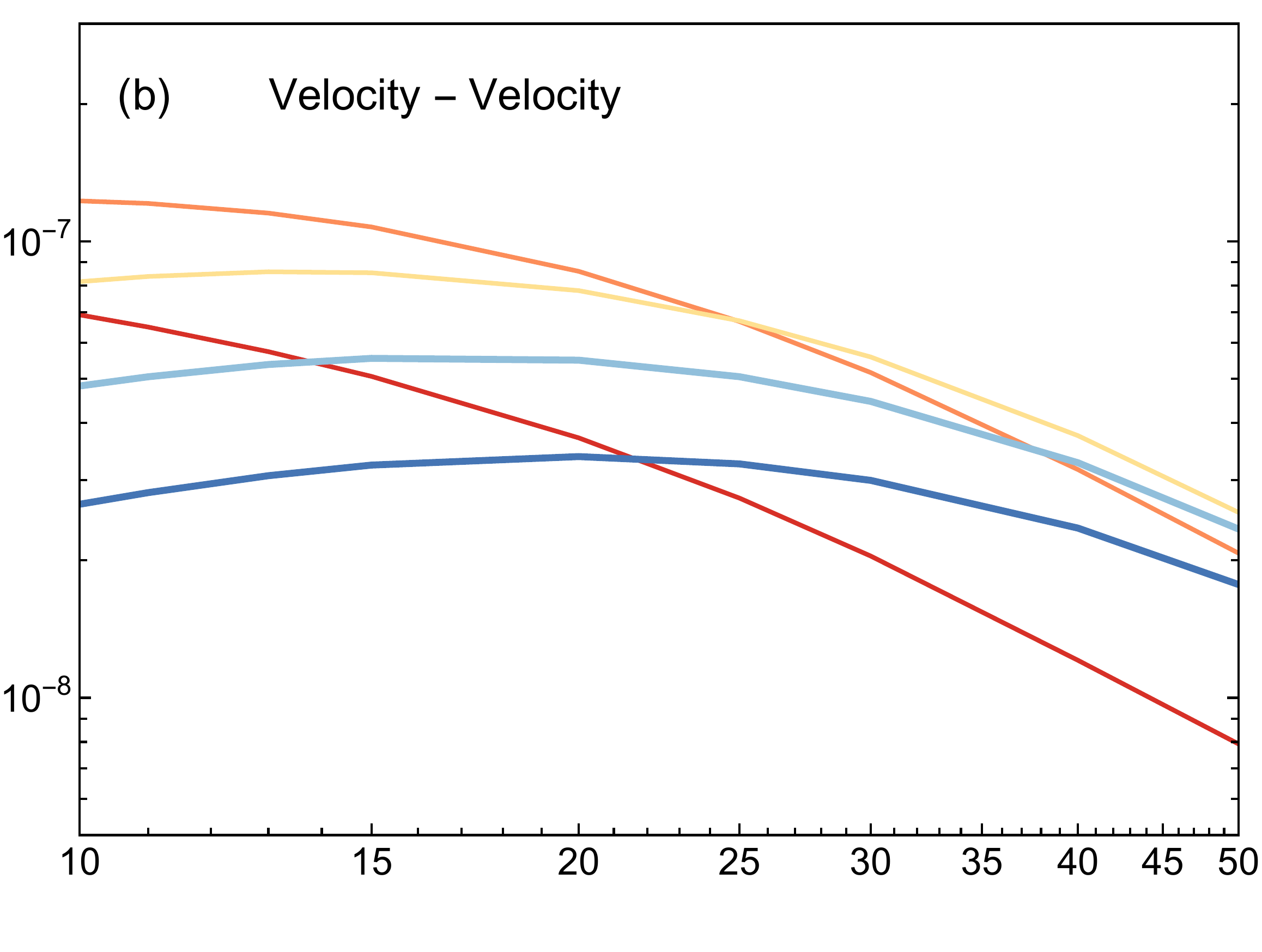}
\end{subfigure}

\vspace{-0.3cm}

\begin{subfigure}[b]{0.49\textwidth}
\centering\includegraphics[height=6.2cm]{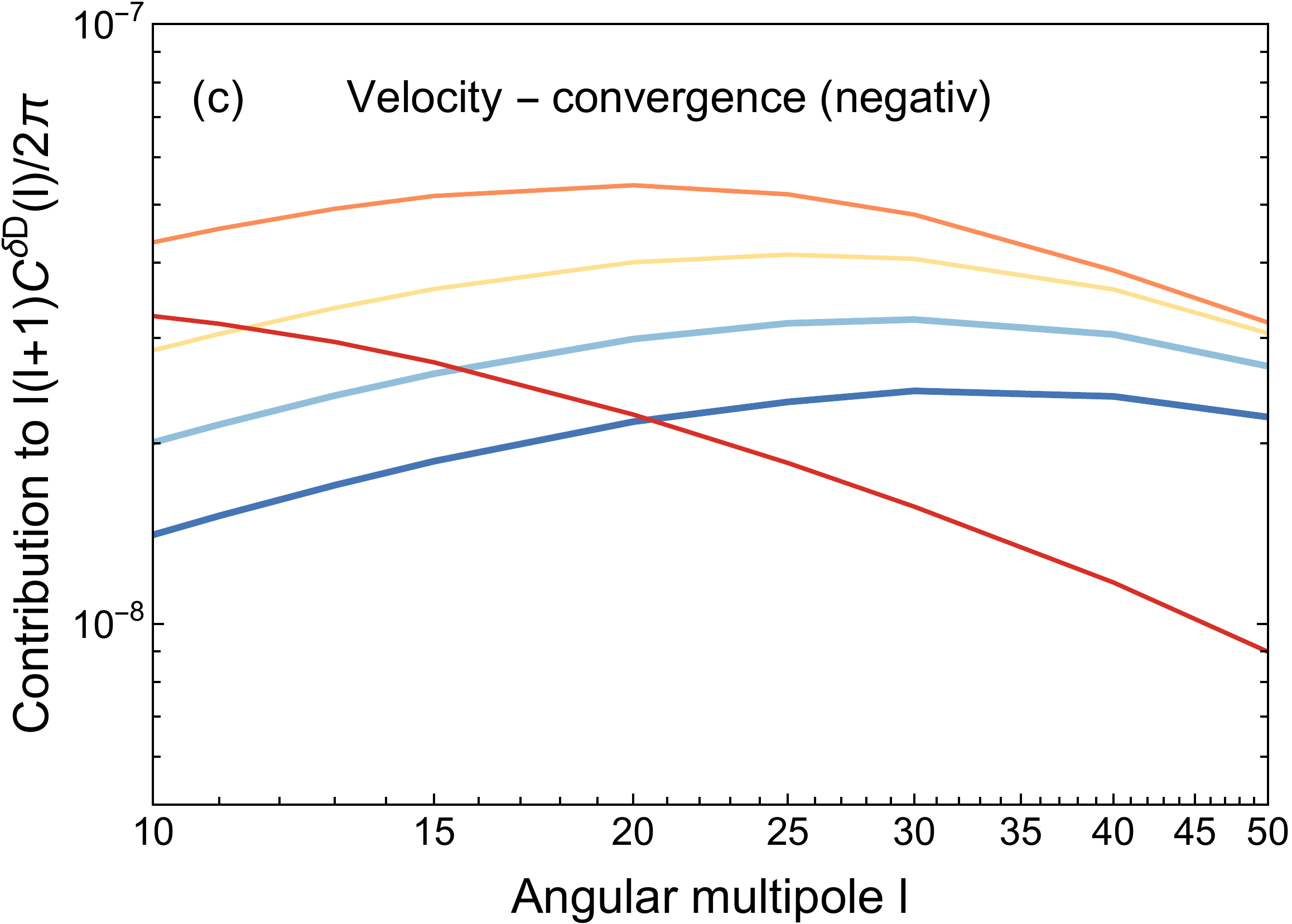}
\end{subfigure}
\begin{subfigure}[b]{0.49\textwidth}
\centering\includegraphics[height=6.2cm]{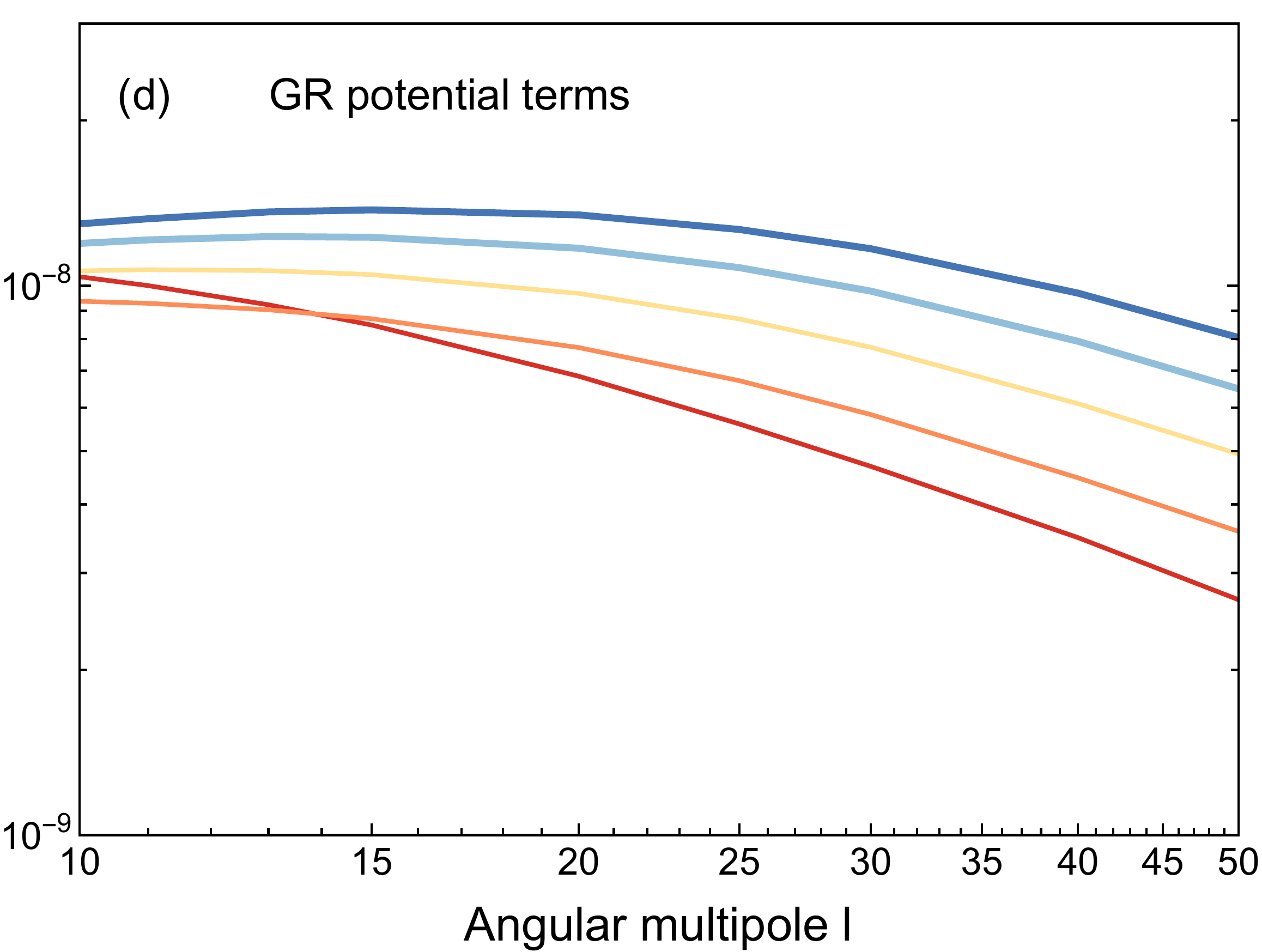}
\end{subfigure}

\caption{Contribution to the total magnification angular power spectrum of the first five redshift bins of Euclid's weak lensing survey caused by (a) the standard convergence, (b) the velocity-velocity contribution, (c) the velocity-convergence cross term and (d) the additional GR potential terms. Note that for the negative velocity-convergence contribution, we plot the absolute value.} \label{Magnification1}
\end{center}
\end{figure*}

In this section, we numerically evaluate the general relativistic effects for the magnification angular power spectrum (Sec.~\ref{ResultsD}) and the cross angular power spectrum between the magnification and shear E-modes (Sec.~\ref{ResultsE}). For these evaluations, we apply a Euclid-like redshift distribution function~\cite{Amendola:2016saw},
\beeq
 n_{z}(z)=\frac{3}{2z_\ast}\rbr{\frac{z}{z_\ast}}^2\exp(-(z/z_\ast)^{3/2})\,, \label{nEuclid} 
 \eneq
where $z_\ast$ is related to the median $z_m$ as $z_\ast=z_m/1.412$, and $z_m=0.9$. As specified in~\cite{Blanchard:2019oqi}, Euclid's weak lensing survey will be split into 10 equi-populated redshift bins with boundaries $\{0.001,$ $0.42,$ $0.56,$ $0.68,$ $0.79,$ $0.90,$ $1.01,$ $1.15,$ $1.32,$ $1.58,$ $2.50\}$. We will only show results for the lowest five of these redshift bins, as general relativistic effects are insignificant for the higher bins compared to the standard convergence. 
All quantities are evaluated for the angular multipoles $l=10,11,\dots,50$, as $l_\mathrm{min}=10$ is the relevant lower boundary for the Euclid survey~\cite{Blanchard:2019oqi} and the total contribution of general relativistic effects is below 1\% at $l=50$ for all bins. 

We apply the relations for the growth functions given in Sec.~\ref{Section:Perturbations}, along with the matter power spectrum $P_{m,0}$ at $a_o=1$ evaluated by  \texttt{CLASS}~\cite{Blas:2011rf}. We assume $H_0=67.4\,\mbox{km}/\mbox{s}/\mbox{Mpc}$ for the Hubble constant, $\Omega_bh^2=0.0224$ and $\Omega_{{cdm}}h^2=0.12$ for the baryonic and dark matter density, $n_s=0.966$ for the scalar spectral index and $A_s=2.1\times 10^{-9}$ for the scalar amplitude at the pivot scale $k_0=0.05/\mbox{Mpc}$, consistent with the Planck 2018 results~\cite{Aghanim:2018eyx}. 

\subsection{Angular power spectrum of the magnification} \label{ResultsD}

\begin{figure*} 

\begin{center}
\begin{subfigure}[b]{1\textwidth}
\includegraphics[width=0.7\linewidth]{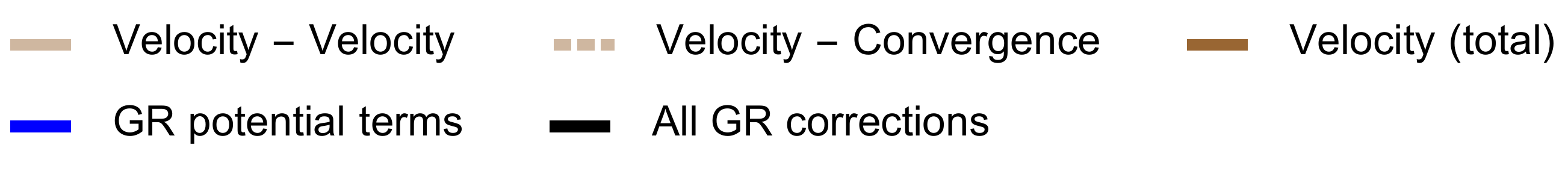} 
\end{subfigure}

\hfill

\begin{subfigure}[b]{0.49\textwidth}
\centering \includegraphics[height=6.25cm]{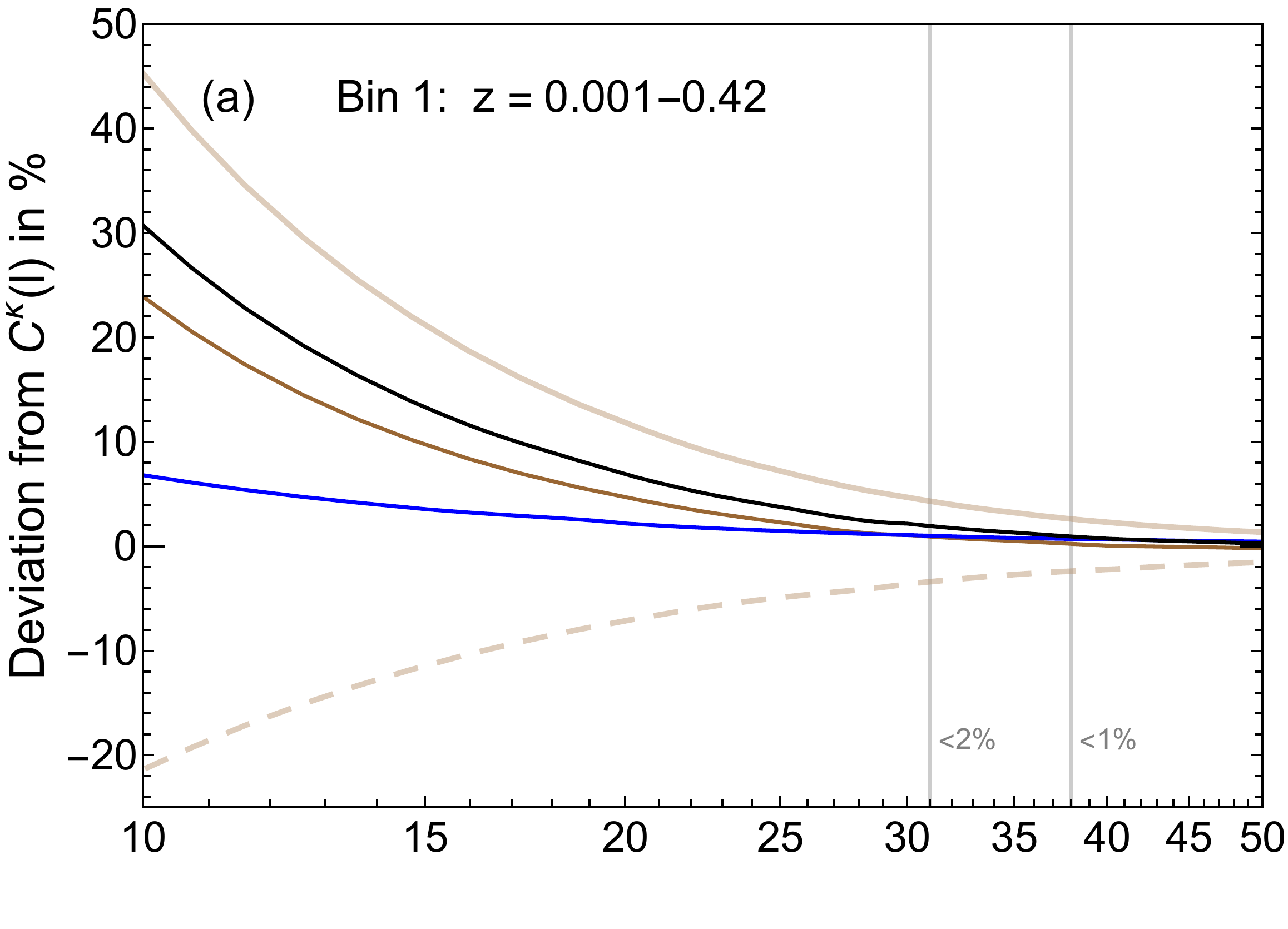}
\end{subfigure}
\begin{subfigure}[b]{0.49\textwidth}
\centering \includegraphics[height=6.25cm]{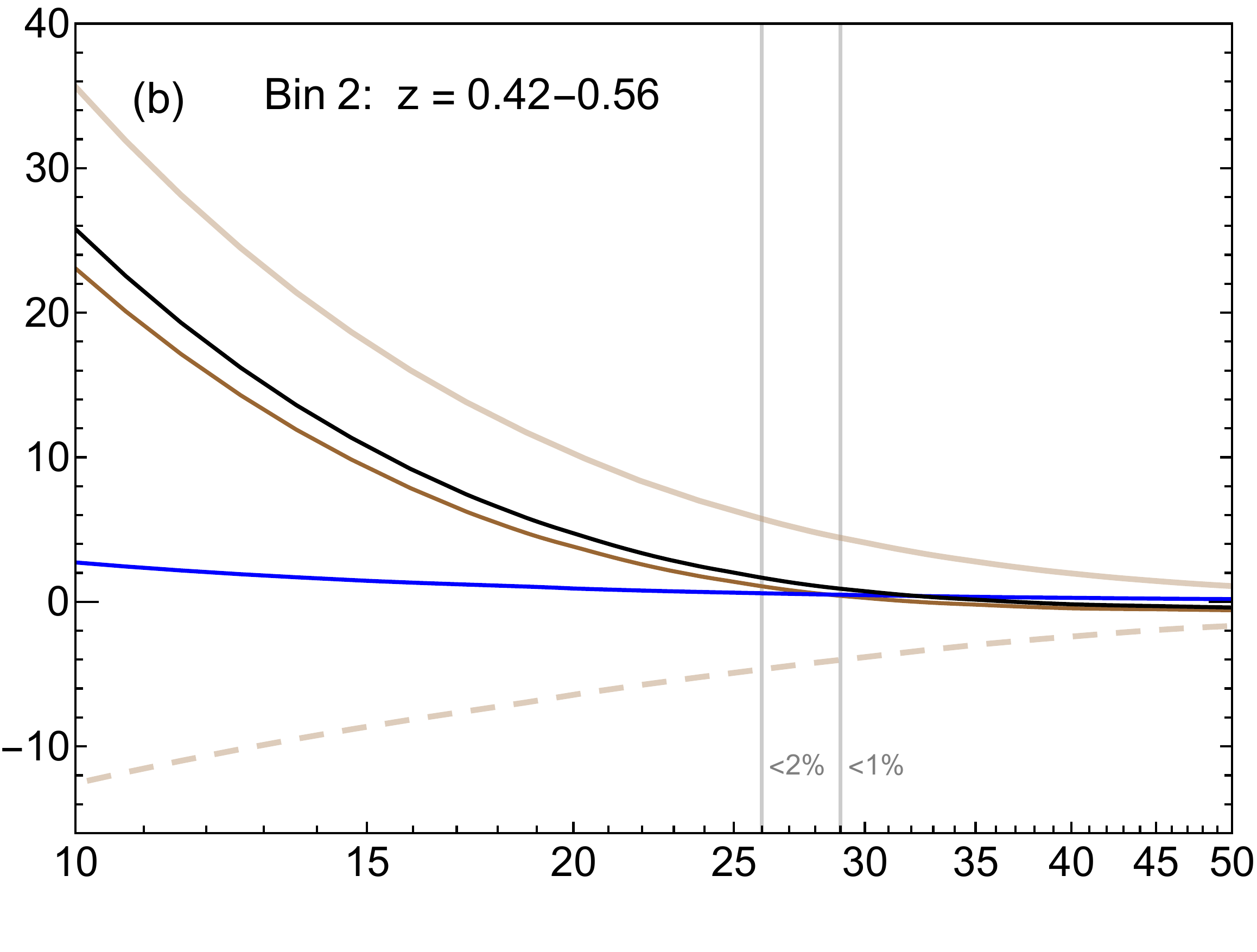}
\end{subfigure}

\vspace{-0.3cm}

\begin{subfigure}[b]{0.49\textwidth}
\centering\includegraphics[height=6.25cm]{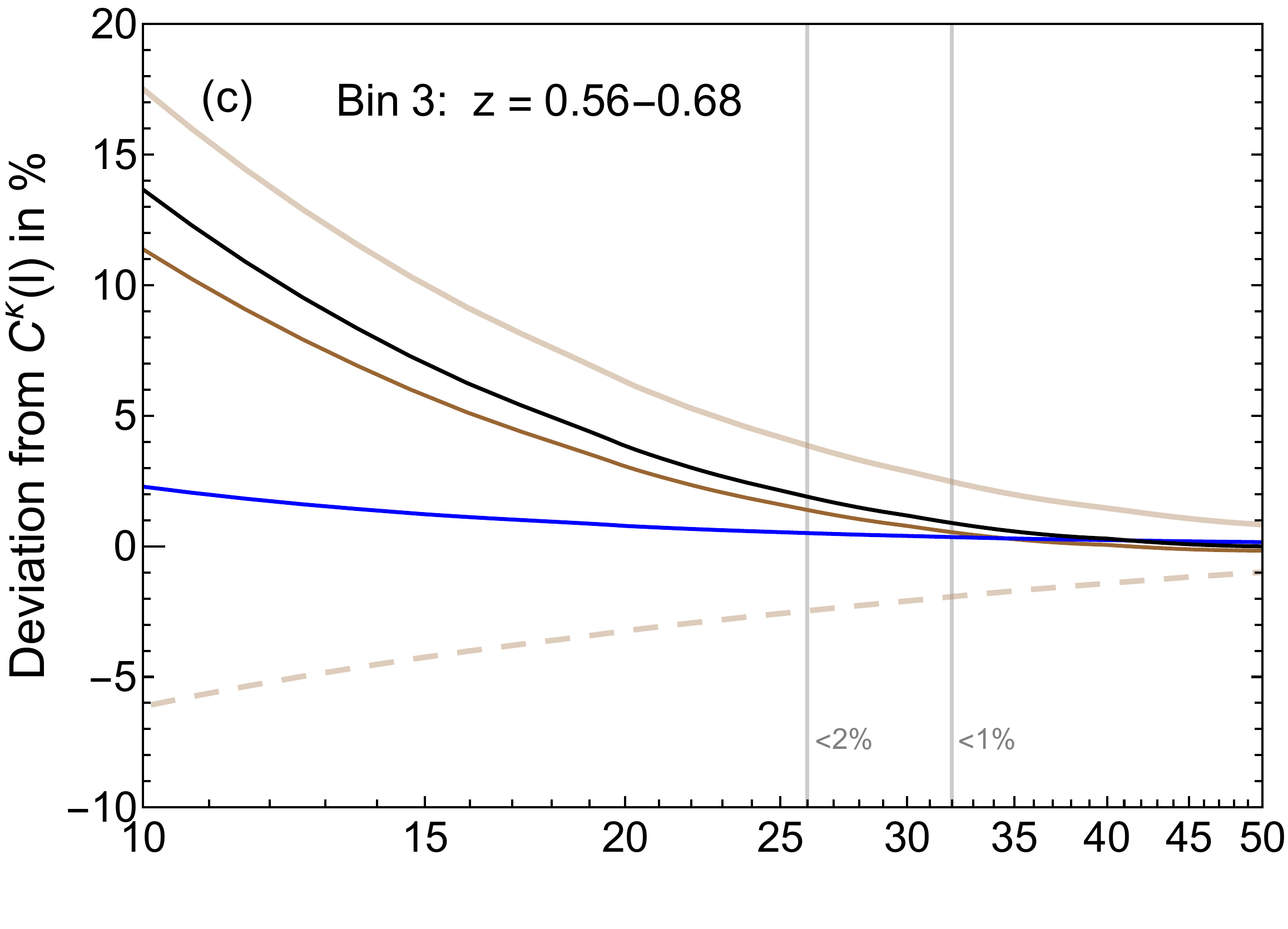}
\end{subfigure}
\begin{subfigure}[b]{0.49\textwidth}
\centering\includegraphics[height=6.25cm]{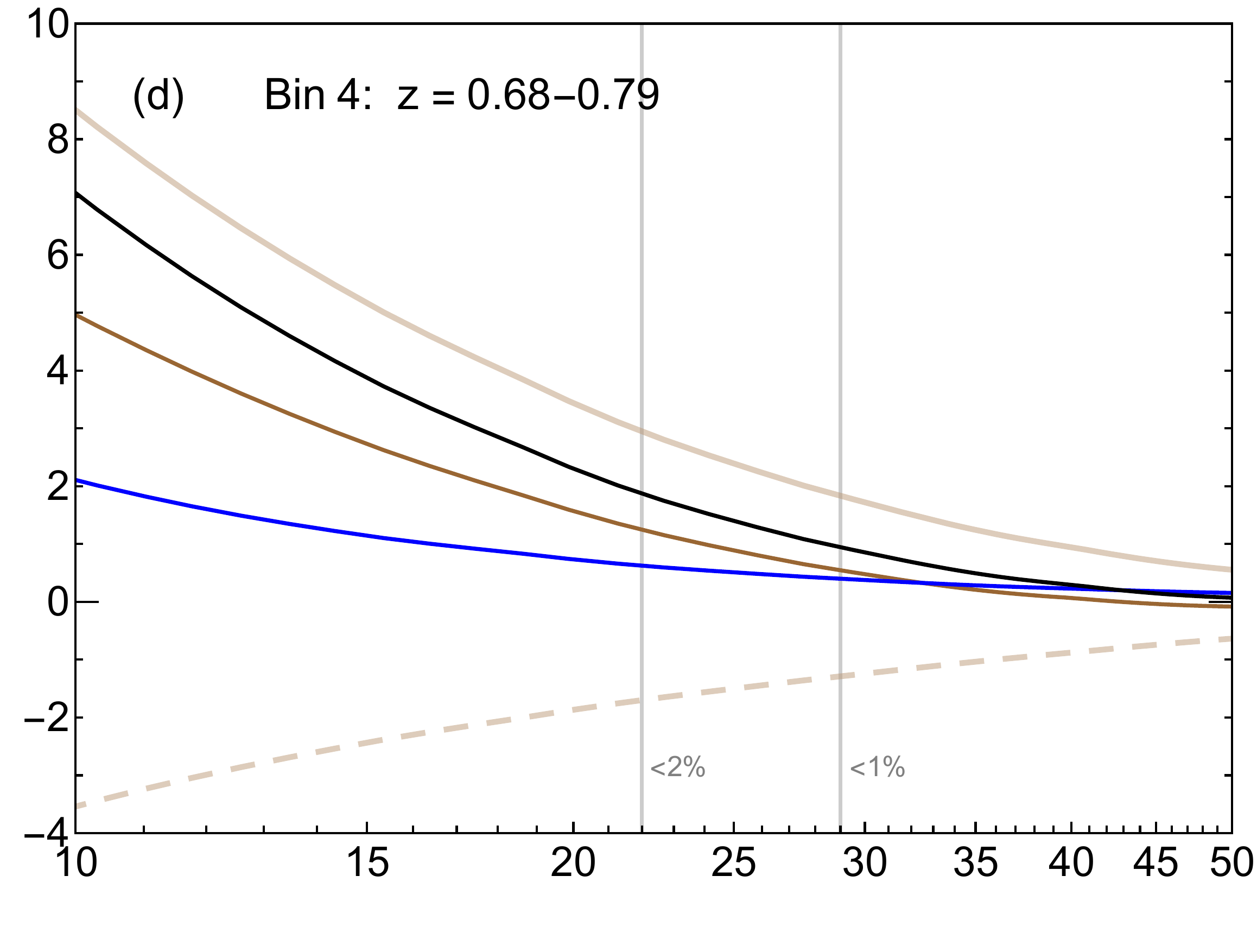}
\end{subfigure}

\vspace{-0.3cm}

\begin{subfigure}[b]{0.49\textwidth}
\centering\includegraphics[height=6.25cm]{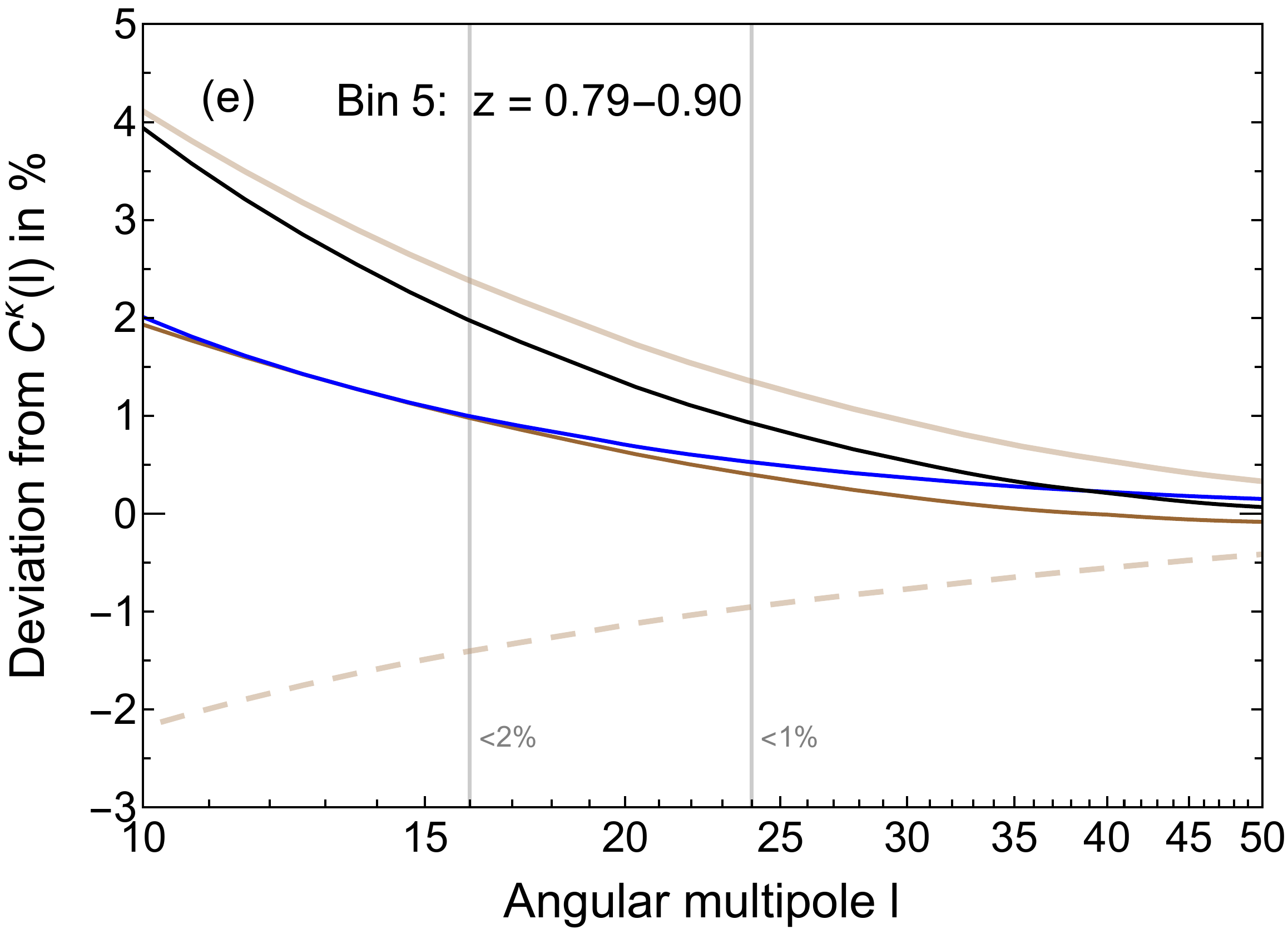}
\end{subfigure}
\begin{subfigure}[b]{0.49\textwidth}
\centering\includegraphics[height=6.25cm]{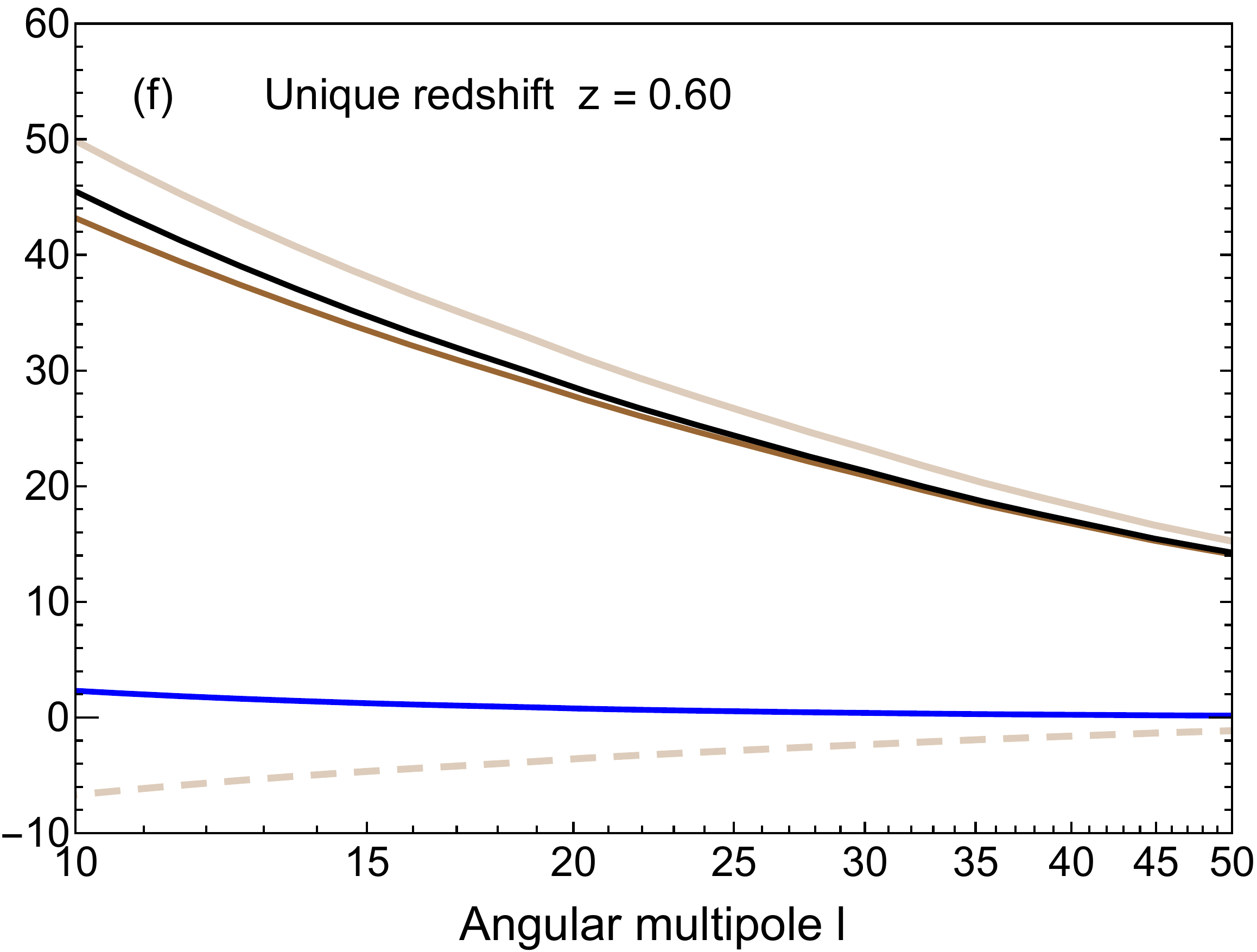} 
\end{subfigure}

\caption{Impact of different relativistic corrections on the magnification angular power spectrum for the first five bins of Euclid's weak lensing survey (panel (a)-(e)), and for a unique source redshift at $z=0.6$ (panel (f)). For panel (a)-(e), gray vertical lines mark the first $l$ for which the total deviation caused by general relativistic effects (black curve) drops below 2\% and 1\%. }\label{Magnification2}
\end{center}

\vspace{1cm} 
\end{figure*}

In Fig.~\ref{Magnification1}, we show the different contributions to the total magnification angular power spectrum for the first five Euclid redshift bins. The contributions of the standard convergence and the GR potential terms are shown in panels~\href{Magnification1}{(a)} and~\href{Magnification1}{(d)}, respectively. The Doppler magnification is split into its two different ways by which it affects the total magnification angular power spectrum: panel~\href{Magnification1}{(b)} shows the velocity-velocity contribution, i.e.~the angular power spectrum of the Doppler magnification. While this contribution is in principle larger for low redshifts due to the prefactor $\As=1-(\mathcal H_s\rs)^{-1}$, the width of the redshift bin is also highly important. In particular, the large width of the first bin leads to a drastic reduction of the signal. Panel~\href{Magnification1}{(c)} shows the second non-negligible contribution of the Doppler magnification, which is the velocity-convergence cross term (i.e., the cross angular power spectrum between the Doppler magnification and standard convergence). This contribution, again determined by both the width and depth of the redshift bin, has a negative value and reduces the overall velocity signal.

In Fig.~\ref{Magnification2}, we see that the velocity-convergence cross term indeed significantly reduces the overall velocity contribution, making it even negative at $l\gtrsim 35$. 
{For bin 4 the overall velocity contribution is already below 5\%, and for bin 5 below 2\% at all observable angular multipoles. Since the width of the redshift bin washes out the velocity contribution, thinner bins would increase the signal. In panel~\href{Magnification2}{(f)}, we provide a comparison to the idealized case of an infinitely thin redshift bin at $z=0.6$. Here, the Doppler magnification still leads to a correction of 15\% at $l=50$ compared to the standard convergence. Nevertheless, even in this idealized case we see that the reduction from the Doppler-convergence cross term is still significant (the amplitude of the term $\propto S_l^\kappa(k)S_l^v(k)$ is about {$13.4\%$} of the $\propto S_l^v(k)^2$ term at $l=10$, and {$7.5\%$} at $l=50$). Hence, this term should not be neglected when comparing the magnification angular power spectrum $C^{\delta D}(l)$ to the shear E-mode angular power spectrum $C^{\cale}(l)$, contrary to~\cite{Amendola:2016saw} where only the standard term $\propto S_l^\kappa(k)^2$ and the velocity-velocity term $\propto S_l^v(k)^2$ had been studied. }  

\begin{figure} 
\begin{center}

\includegraphics[width=\linewidth]{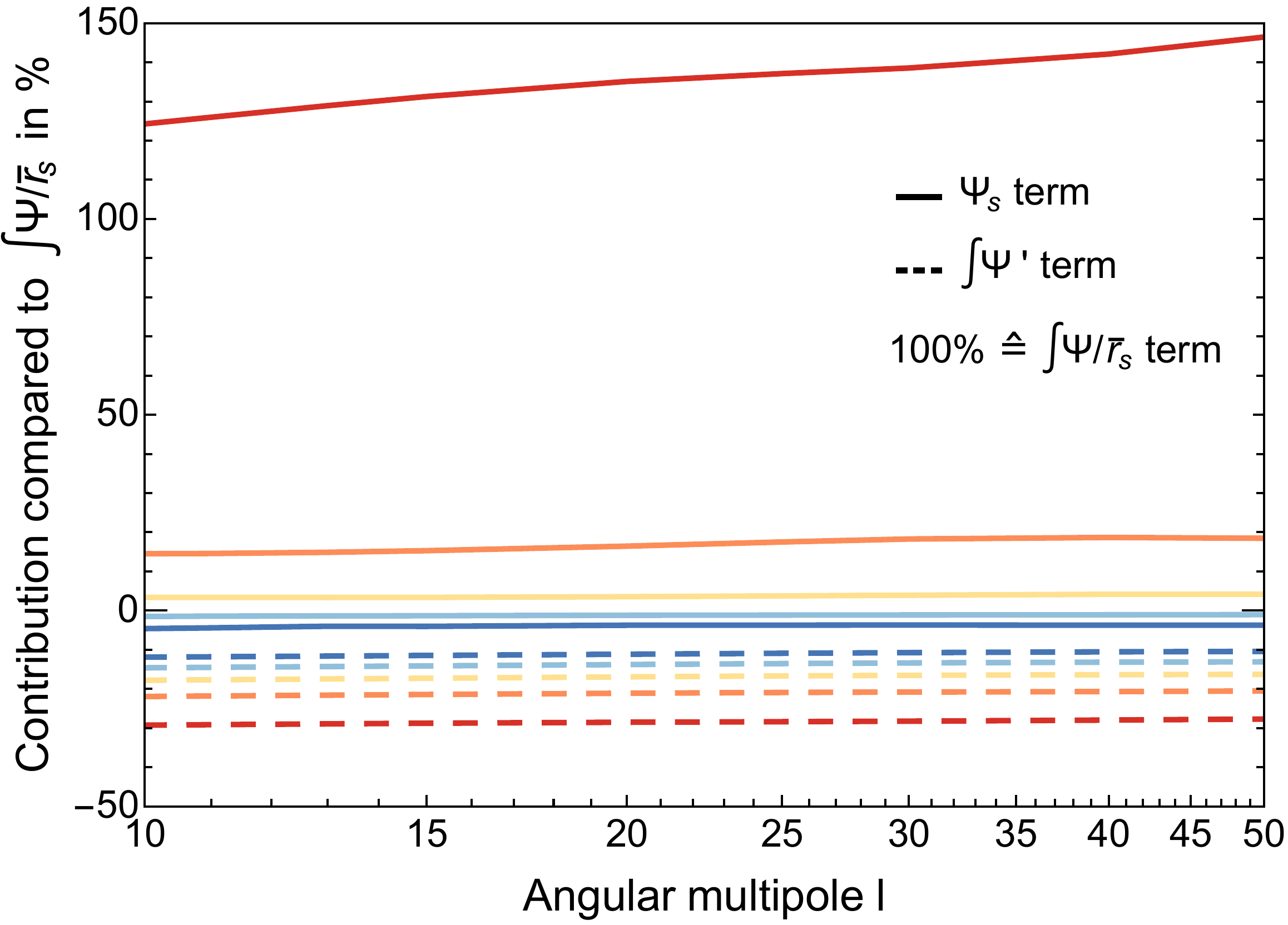}

\caption{Contribution of the $\Psi_s$ term (solid) and the $\uint\Psi'$ term (dashed lines), in comparison to the $\uint\Psi/\rs$ term at each bin. The different colors correspond to the five redshift bins as in Fig.~\ref{Magnification1}. } \label{ContPsi}
\end{center}
\end{figure}

For the GR potential terms (blue lines in Fig.~\ref{Magnification2}), the fractional contribution to $C^{\delta D}(l)$ does not vary significantly between the redshift bins 2--5: It is at 2--3\% for $l_\mathrm{min}=10$, and quickly falls below 1\% at $l=20$. Only for bin 1, the contribution is larger, leading to a deviation of about 7\% at $l_\mathrm{min}=10$. The reason is the behavior of the different GR potential terms at low and high redshifts. As seen in Eq.~\eqref{deltaDdiffcont}, $\delta D^{p}$ consists of a $\Psi_s$ term (with the potential evaluated at the source position), a $\uint\Psi/\rs$ term and a $\uint\Psi'$ term (with the potential and, respectively, its time variation integrated along the line of sight). In Fig.~\ref{ContPsi}, the contributions of the $\Psi_s$ and $\uint\Psi'$ terms to the magnification angular power spectrum are compared to the contribution of the $\uint\Psi/\rs$ term. We see that for the redshift bins 2--5, the $\uint\Psi/\rs$ term indeed constitutes the major GR potential contribution. For bin 1, however, the $\Psi_s$ term enhances it by more than $120\%$. This is due to the pre-factor $-(\As+1)=(\mathcal H_s\rs)^{-1}-2$, which is large at low redshift, but turns negative at $z\approx 0.682$ and slowly converges to $-2$.  For the negative contribution of the $\uint\Psi'$ term, we see that it gets smaller for higher redshift bins. This is because $\Psi'$ is equal to zero in a matter-only universe. Thus, the integral involving $\Psi'$ gets its major contribution from low $z$, and does not grow with a larger source redshift as much as the $\uint\Psi/\rs$ term.  

While the contribution of GR potential terms alone would be a rather low correction, it enhances the signal arising from the velocity. Indeed, as seen in panel~\href{Magnification2}{(e)}, the contribution of the GR potential terms is similarly large as the contribution of the Doppler magnification in the depicted range of $l$, although the combined signal is low compared to cosmic variance in this regime. Moreover, panel~\href{Magnification2}{(b)} shows that even for bin~2 they still enhance the signal by more than 10\%{, partly counteracting the reduction from the convergence-velocity cross term.} 

In Fig.~\ref{Magnification1}--\ref{Magnification2}, we only show the results for the lower five Euclid redshift bins. For the higher bins, the additional GR potential terms lead to a practically identical fractional contribution as for bin 5, i.e., it is at 2\% at $l_\mathrm{min}=10$ and below 1\% at $l=20$. The Doppler magnification, diminished at high redshift due to the pre-factor $\As=1-(\mathcal H_s\rs)^{-1}$, is even smaller. Hence, general relativistic corrections play an insignificant role for the upper half of redshift bins.

\subsection{Cross angular power spectrum between the magnification and shear E-modes} \label{ResultsE}

\begin{figure*} 
\begin{center}

\begin{subfigure}[b]{1\textwidth}
\includegraphics[width=0.7\linewidth]{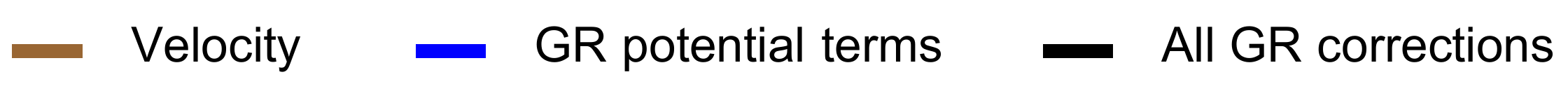} 
\end{subfigure}

\begin{subfigure}[b]{0.49\textwidth}
\centering\includegraphics[height=6.2cm]{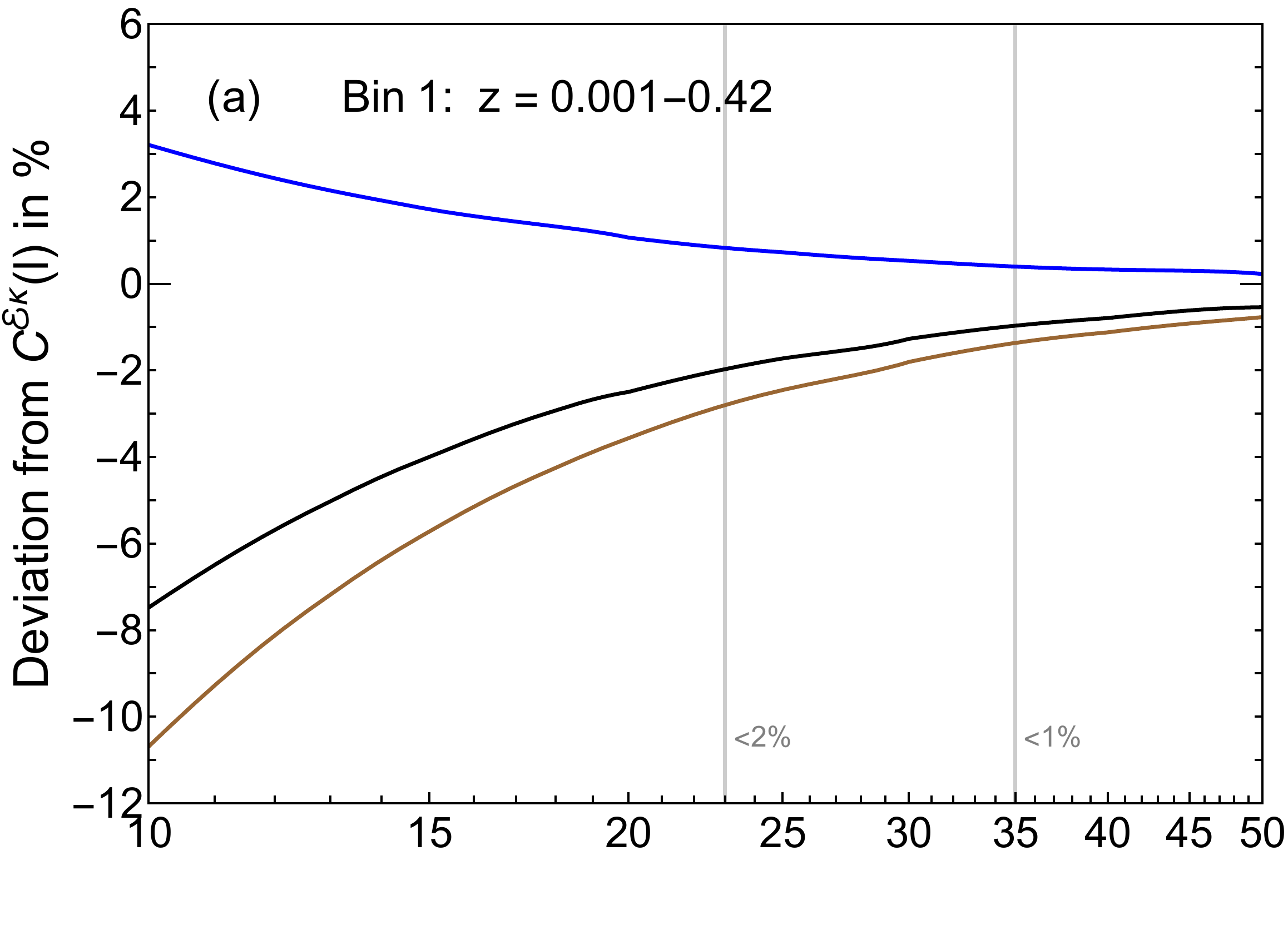}
\end{subfigure}
\begin{subfigure}[b]{0.49\textwidth}
\centering\includegraphics[height=6.2cm]{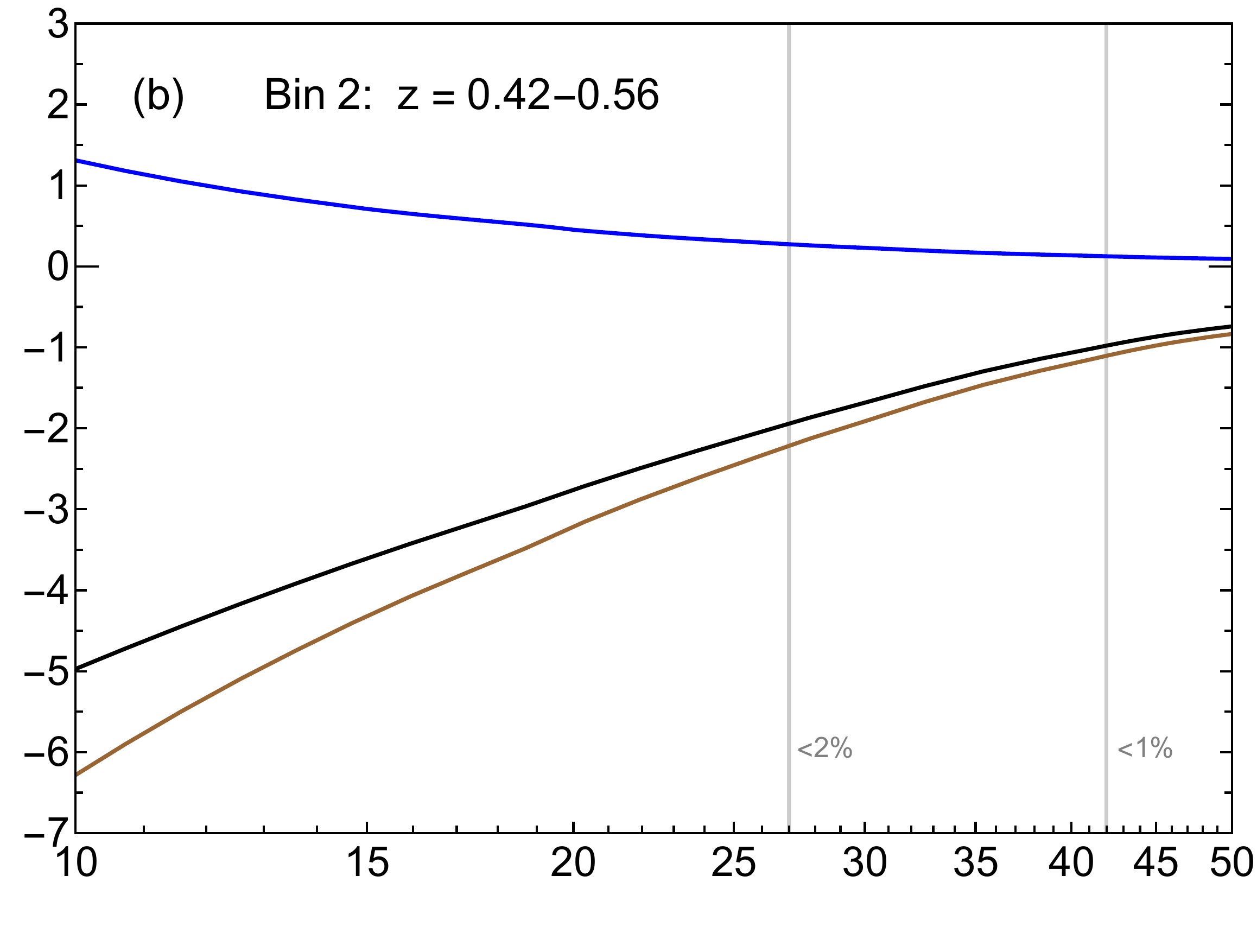}
\end{subfigure}

\vspace{-0.3cm}

\begin{subfigure}[b]{0.49\textwidth}
\centering\includegraphics[height=6.2cm]{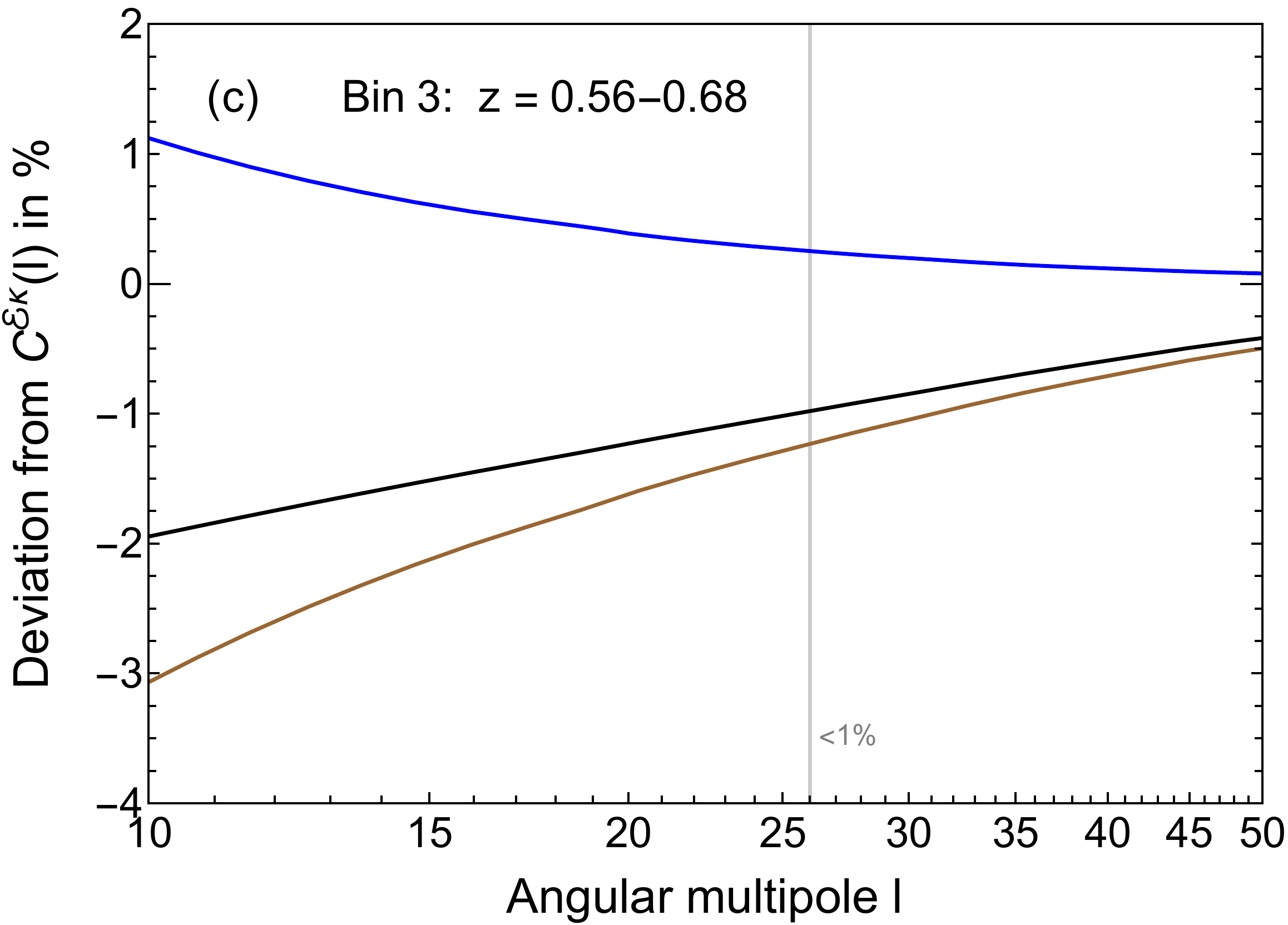}
\end{subfigure}
\begin{subfigure}[b]{0.49\textwidth}
\centering\includegraphics[height=6.2cm]{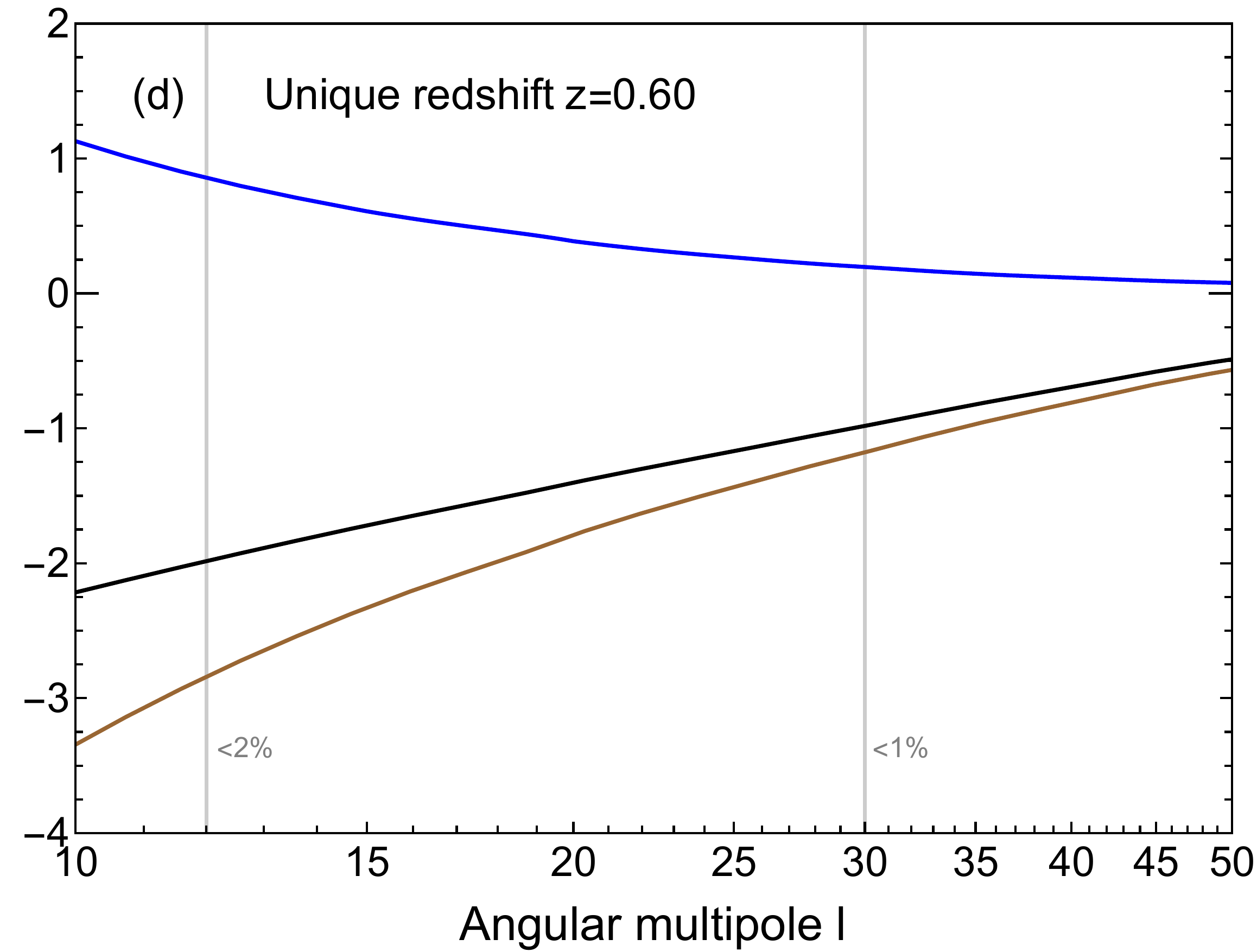}
\end{subfigure}

\caption{Impact of velocity terms (i.e., the Doppler magnification) and additional GR potential terms on the cross angular power spectrum of the magnification and shear E-modes, for the first three Euclid redshift bins and a unique source redshift $z=0.60$. Gray vertical lines mark the first $l$ for which the absolute value of the total deviation caused by general relativistic effects (black curve) drops below 2\% and 1\% (unless it is already below that value at $l_\mathrm{min}=10$).} \label{CrossME}
\end{center}
\end{figure*}

We have demonstrated that the Doppler magnification alters the magnification angular power spectrum not only via its auto correlation, but also via the velocity-convergence cross term. Likewise, the Doppler magnification along with the GR potential terms also affect the cross angular power spectrum $C^{\delta D\cale}(l)$, as illustrated in Fig.~\ref{CrossME}. {As explained at the end of Sec.~\ref{Section:WLO}, measuring the corresponding terms $\propto S_l^\kappa(k)(S_l^v(k)+S_l^p(k))$ in the cross spectrum $C^{\delta D\cale}(l)$ by comparing to the shear auto spectrum $C^\cale(l)$ would provide us a way to isolate the pure Doppler lensing term $\propto S^v_l(k)^2$ in the magnification auto spectrum $C^\kappa(l)$. 
However, in the cross power spectrum $C^{\delta D\cale}(l)$ the contribution of the terms~$\propto S_l^\kappa(k)(S_l^v(k)+S_l^p(k))$ is reduced by a factor of~2 compared to their contribution to the auto power spectrum $C^{\delta D}(l)$, further complicating their measurement. For bin~3, the resulting signal is already below 2\% at all scales. 
At higher redshifts, the velocity contribution will eventually become irrelevant. The total contribution is then fully determined by the GR potential terms, leading to a correction of  1\% or less.}

{In panel~\href{Magnification2}{4.(d)}, we provide a comparison for an infinitely thin redshift bin at $z=0.60$, showing that the signal of the terms~$\propto S_l^\kappa(k)(S_l^v(k)+S_l^p(k))$ is still low in this idealized case, making an individual measurement unlikely.}



\section{Summary and conclusion} \label{Section:Conclusion}

\begin{figure} 
\begin{center}

\includegraphics[width=\linewidth]{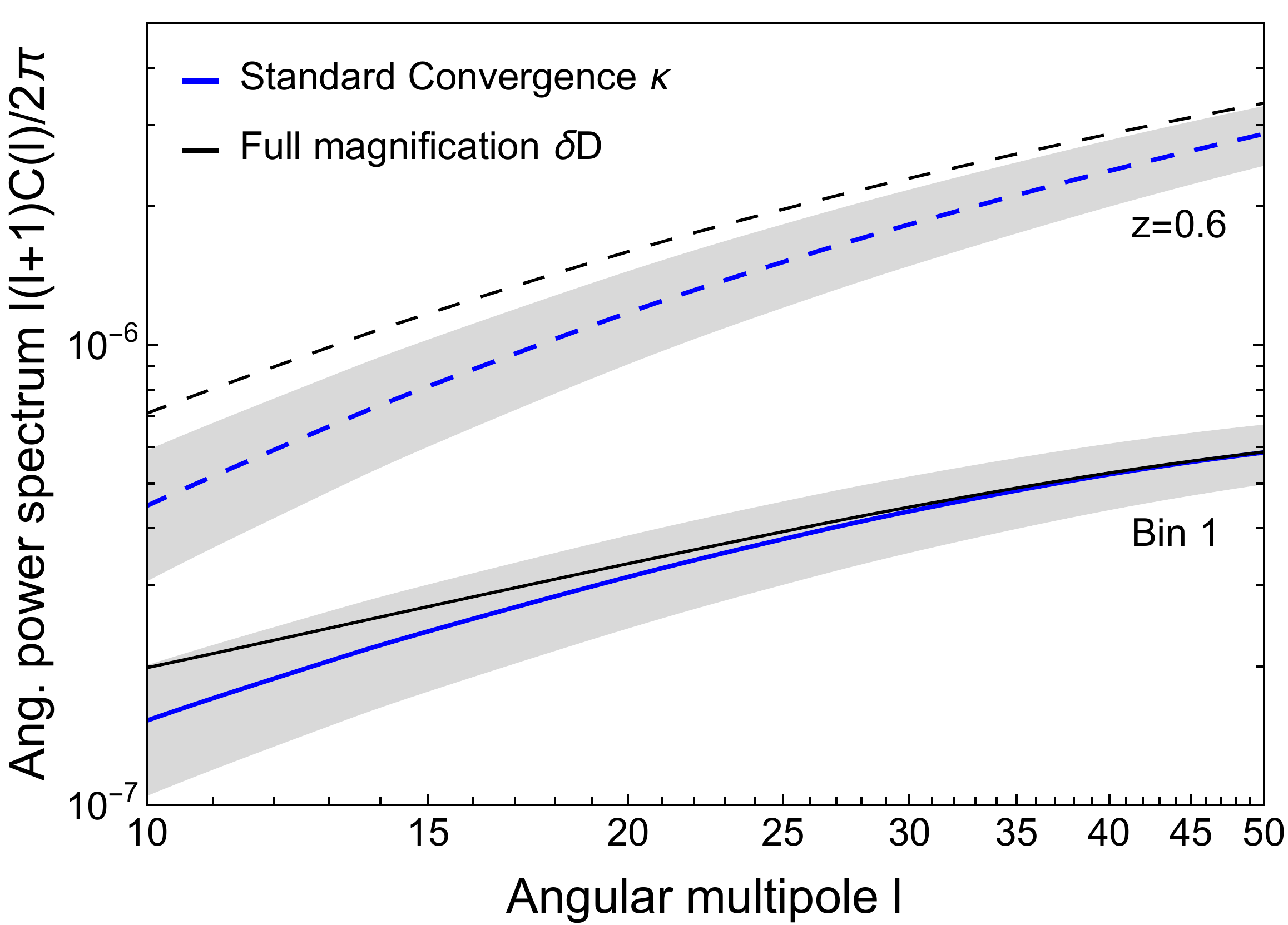}

\caption{Angular power spectra of the standard convergence $\kappa$ and the full magnification observable $\delta D$ for the lowest Euclid bin ($z=0.001-0.42$; solid) and a unique source redshift at $z=0.60$ (dashed). The gray regions indicate cosmic variance.} \label{CosVar}
\end{center}
\end{figure}

{In this paper, we have numerically investigated the impact of all general relativistic effects on the weak lensing power spectra on large scales measurable in a survey like Euclid.}
We have shown that the width of the redshift bins along with the negative velocity-convergence cross term drastically reduce the Doppler magnification signal. 
For the magnification angular power spectrum, the total impact of general relativistic effects is still large at $l_\mathrm{min}=10$ (4--30\%, depending on the considered redshift bin), but falls off fast and is below 1\% at $l=38$ even for the lowest redshift bin. For the shear-magnification cross angular power spectrum, general relativistic effects lead to a rather low correction already at $l_\mathrm{min}=10$ (with an absolute value of 7.5\% for the lowest bin), 
and is below $1\%$ at $l=42$ for all bins. 
Taking into account the large cosmic variance at large angular scales, $\Delta C(l)/C(l)=\sqrt{2/(2l+1)}$, this could pose a serious challenge in measuring the relativistic corrections. For further illustration, we show in Fig.~\ref{CosVar} the deviation of the full observable $C^{\delta D}(l)$ from the standard prediction $C^\kappa(l)$ for the lowest Euclid redshift bin (where the deviation is larger compared to the other bins) as well as a unique source redshift $z=0.6$. For the latter case, the impact of general relativistic effects is above cosmic variance. However, for Bin~1, it reaches a comparable level only at the lowest multipole $l_\mathrm{min}=10$.

One obvious solution to enhance the Doppler magnification signal would be to reduce the width of the redshift. However, several obstacles for doing so need to be pointed out. Indeed, in~\cite{Bacon:2014uja} only averages over galaxy pairs separated by no more than $\Delta z=0.02$ are applied, but the direct measurement of the Doppler magnification via galaxy sizes is only possible up to $z\approx 0.3$ (or $z\approx 0.5$ when using the dipole of the cross-correlation between galaxy sizes and number counts~\cite{Bonvin:2016dze}) as the standard convergence becomes significant for higher redshifts. For the method proposed in~\cite{Amendola:2016saw} with respect to the Euclid survey, both the magnification and cosmic shear need to be measured in order to extract general relativistic effects by applying Eq.~\eqref{relation}.
 This assumes that the same source redshift distribution is applied for both measurements. Hence, reducing the width of the magnification measurement requires to do the same for the cosmic shear measurement. This is however challenging due to the various systematic uncertainties in cosmic shear measurements, including photometric redshifts as well as difficulties in shape measurements (see e.g.~\cite{Mandelbaum:2017jpr}). Moreover, measurements of magnification itself are also limited by systematics, and in particular require sufficient knowledge of intrinsic sizes and magnitudes~\cite{Alsing:2014fya}. {We also emphasize that even for thinner bins the velocity-convergence cross term $\propto S_l^\kappa(k)S_l^v(k)$ still needs to be taken into account, as illustrated by the idealized case of an infinitely thin redshift bin at $z=0.60$ in Fig.~\href{Magnification2}{2.(f)}.} 

Despite the difficulties, a reduced width for the redshift distribution in weak lensing measurements certainly seems to be the best approach for increasing the observability of the Doppler magnification {by measuring both magnification and shear as proposed in~\cite{Amendola:2016saw}}. Another theoretical possibility would be to extend to an all-sky survey, measuring even the lowest $l<10$. As seen especially in Fig.~\ref{Magnification2}, the impact of general relativistic effects increases fast when approaching $l_\mathrm{min}=10$, and their even larger contribution at $l<10$ could compensate for the high variance at such large angular scales. However, such a weak lensing survey will not be realized in the near future. {Quantifying detectability of the relativistic effects in the lensing power spectra will require a careful Fisher matrix analysis in a given survey geometry, which is beyond our current scope to investigate the impact of the relativistic effects on the weak lensing power spectra.}

Finally, we want to point out that general relativistic effects should not only be seen as corrections to standard effects, but also as an important cosmological probe themselves. In particular, as pointed out in~\cite{Amendola:2016saw}, the standard lensing convergence is determined by the sum of the Bardeen potentials $\Phi+\Psi$, but the additional general relativistic corrections break the degeneracy between them. While we did not distinguish between these two potentials in this work, as general relativity with no anisotropic stress component predicts $\Phi=\Psi$, modified gravity theories can lead to deviations from this assumption. Therefore, it can lead to a modified signal at low $l$.

We conclude that, while measuring general relativistic effects in weak lensing angular power spectra is difficult, it is certainly not impossible. These general relativistic effects at large angular scales provide a test for theories of gravity, and therefore deserve the attention from both the theoretical and observational community.

\begin{acknowledgments}
We thank Fulvio Scaccabarozzi and Giuseppe Fanizza for useful discussions. We acknowledge support by the Swiss National Science Foundation (SNF CRSII5\_173716), and J.Y.~is further supported by a Consolidator Grant of the European Research Council (ERC-2015-CoG grant 680886).
\end{acknowledgments}

\appendix

\section{Vector calculus identities in spherical coordinates} \label{Appendix:VectorCalc}

Observed source positions on the sky are usually described by spherical coordinates -- a radial coordinate $\bar r(z)$ associated to the observed redshift $z$, and observed angular coordinates $(\theta,\phi)$. Here, we review some vector calculus identities in spherical coordinates that are fundamental for this work. First of all, the line-of-sight direction $n^\alpha$ specified by the observed angles $\theta$ and $\phi$ is given by
\begin{align}
\mathbf n=n^\alpha(\theta,\phi)=\begin{pmatrix}\sin\theta\cos\phi\\\sin\theta\sin\phi\\\cos\theta\end{pmatrix}\,.
\end{align}
Introducing two vectors orthonormal to $n^\alpha$,
\begin{align}
\theta^\alpha(\theta,\phi)=\begin{pmatrix}\cos\theta\cos\phi\\\cos\theta\sin\phi\\-\sin\theta\end{pmatrix}\,,\qquad\phi^\alpha(\theta,\phi)=\begin{pmatrix}-\sin\phi\\\cos\phi\\0\end{pmatrix}\,,
\end{align}
we can write the gradient as
\beeq
\frac{\partial}{\partial x^\alpha}=n_\alpha \partial_{\bar r}+\frac{1}{\bar r}\habla_\alpha\,,\qquad \habla_\alpha\equiv\theta_\alpha \partial_\theta +\phi_\alpha\sin^{-1}\theta\, \partial_\phi\,, \label{gradient}
\eneq
where we refer to $\habla_\alpha$ as the angular gradient. Furthermore, we define the angular Laplacian operator,
\beeq
\habla^2=\habla_\alpha\habla^\alpha={\partial_\theta^2}+{\cot\theta}{\partial_\theta}+\sin^{-2}\theta\,{\partial_\phi^2}\,.
\eneq
Using the angular derivatives of the basis vectors,
\begin{align}
&\partial_\theta \boldsymbol n=\boldsymbol \theta \,,\quad\partial_\theta\boldsymbol \theta=-\boldsymbol n\,,\quad \partial_\theta\boldsymbol \phi=0\,, \quad \sin^{-1}\theta\,\partial_\phi\mathbf n=\mathbf \phi\,, \nnn
&\sin^{-1}\theta\,\partial_\phi\boldsymbol \theta=\cot\theta\,\boldsymbol \phi\,,\quad \sin^{-1}\theta\,\partial_\phi\boldsymbol \phi=-\boldsymbol n-\cot\theta\,\boldsymbol \theta\,, \label{basisderiv}
\end{align}
we can derive the following relations that are used for the calculations in this work. First of all, applying the angular gradient and angular Laplacian to $n^\alpha$ yields
\beeq
\habla_\beta n^\alpha=\theta_\beta\theta^\alpha+\phi_\beta\phi^\alpha=-n_\beta n^\alpha+\delta^\alpha_\beta\,, \qquad \habla^2 \boldsymbol n=-2\boldsymbol n\,. \label{gradientLaplacen}
\eneq

We emphasize that the observed angular direction $\boldsymbol n$ and the two orthonormal directions $\boldsymbol{\theta}$ and $\boldsymbol{\phi}$ are defined in the observer rest frame only. Thus, the definitions presented here are independent of any FRW coordinates in the space-time manifold.

\section{Contribution of different $\mathbf k$-vectors to the angular power spectrum} \label{Appendix:kdirection}

Given a spin-0 quantity $A(\mathbf{ n})$, such as the magnification or the spin-raised and -lowered shear components, the resulting angular power spectrum is given by
\begin{align}
C^A(l)=&\angbr{a^A_{lm}a^{A\ast}_{lm}}=\iint \frac{\mathrm d^3 k}{(2\pi)^3}\frac{\mathrm d^3 k'}{(2\pi)^3}\angbr{a^A_{lm}(\mathbf k)a^{A\ast}_{lm}(\mathbf k')} \nnn
=&\int\frac{\mathrm d^3 k}{(2\pi)^3}\left\vert \tilde a^A_{lm}(\mathbf k)\right\vert^2P_\zeta(k)\,, \label{D2}
\end{align}
where $a_{lm}^A(\mathbf k)$ is defined in Eq.~\eqref{alm}, and we additionally introduced $\tilde a^A_{lm}(\mathbf k)$ as $a^A_{lm}(\mathbf k)\equiv\tilde a^A_{lm}(\mathbf k)\zeta(\mathbf k)$. Thus, $C^A(l)$ needs to be calculated from the contribution of all $\mathbf k$ at all angular directions. However, in our calculations in Sec.~\ref{Section:FullSky} for the shear components and the magnification, we have only calculated $a_{lm}(\mathbf k)$ for a $k$-vector aligned with the $z$-axis, $\mathbf k=k\mathbf{e_z}$, and assumed that this is sufficient to evaluate the resulting angular power spectra. Here, we justify this assumption.  

Let $\mathbf{k}$ be some arbitrary wave-vector, and let $\mathbf{ n}_1$ and $\mathbf{ n}_2$ specify two angular directions. First, note that for any rotation matrix $\mathcal R$, the relation
\beeq
\tilde A(\mathbf k,\mathbf{ n}_1)\tilde A^\ast(\mathbf k,\mathbf{ n}_2)=\tilde A(\mathcal R\mathbf k,\mathcal R\mathbf{ n}_1)\tilde A^\ast(\mathcal R\mathbf k,\mathcal R\mathbf{ n}_2)\,, \label{rotA}
\eneq
applies, where $A(\mathbf k,\mathbf{ n})\equiv \tilde A(\mathbf k,\mathbf{ n})\zeta(\mathbf k)$. This is based on the fact that the dependence of $\tilde A(\mathbf k,\mathbf{ n})$ can be expressed as a dependence on $k$ and the angle $\mathbf k\cdot\mathbf{ n}$, which stays invariant when applying the same rotation to both vectors. Now, defining $\mathbf k\equiv \mathcal R_k \mathbf{e_z}$, $\mathbf{ n}_1\equiv \mathcal R_k\mathbf{ n}'_1$ and $\mathbf{ n}_2\equiv \mathcal R_k\mathbf{ n}'_2$, we obtain
\begin{align}
&\iint\mathrm d\Omega_1\mathrm d\Omega_2\,{\tilde A(\mathbf k,\mathbf{ n}_1)\tilde A^\ast(\mathbf k,\mathbf{ n}_2)}Y_{lm}(\mathbf{ n}_1)Y^\ast_{lm}(\mathbf{ n}_2) \nnn
=&\iint\mathrm d\Omega'_1\mathrm d\Omega'_2\,{\tilde A(k\mathbf{e_z},\mathbf{ n}'_1)\tilde A^\ast( k\mathbf{e_z},\mathbf{ n}'_2)}Y_{lm}(\mathbf{ n}_1)Y^\ast_{lm}(\mathbf{ n}_2) \nnn
=&\left\vert\tilde a^A_{lm}(\mathbf k)\right\vert^2\,,
\end{align} 
where we have performed a change of variables and used Eq.~\eqref{rotA}. By applying the addition theorem of spherical harmonics,
\begin{align}
\sum_mY_{lm}(\mathbf{n}_1)Y^\ast_{lm}(\mathbf{ n}_2)&=\frac{2l+1}{4\pi}P_l(\mathbf{ n}_1\cdot\mathbf{ n}_2) \nnn
&=\sum_mY_{lm}(\mathcal R\mathbf{ n}_1)Y^\ast_{lm}(\mathcal R\mathbf{ n}_2)\,, \label{AdditionTheorem}
\end{align}
it follows that 
\begin{align}
\sum_m\left\vert\tilde a^A_{lm}(\mathbf k)\right\vert^2=\sum_m\left\vert \tilde a^A_{lm}( k\mathbf{e_z})\right\vert^2\,. \label{Rotalm}
\end{align}
Finally, noting that $C^A(l)$ is independent of $m$, we can rewrite Eq.~\eqref{D2} as
\begin{align}
C^A(l)&=\frac{1}{2l+1}\int\frac{\mathrm d^3 k}{(2\pi)^3}\sum_m\left\vert \tilde a^A_{lm}(\mathbf k)\right\vert^2P_\zeta(k) \nnn
&=\frac{1}{2l+1}\int\frac{\mathrm d^3 k}{(2\pi)^3}\sum_m\left\vert \tilde a^A_{lm}(k\mathbf{e_z})\right\vert^2P_\zeta(k)
\,, 
\end{align}
proving that, indeed, we can calculate $C^A(l)$ from the contribution of $k$-vectors aligned with the $z$-axis only. Note that, while $\langle a^A_{lm}a^{A\ast}_{lm}\rangle$ is equal for all $m$ due to statistical isotropy, this does not apply for the contribution of a single $k$-mode. In particular, a $k$-mode aligned with the $z$-axis contributes only to $m=0$, while this would not be true for a general $k$-mode since $a^A_{lm}(\mathbf k)\neq a^A_{lm}(k\mathbf{e_z})$ in general. Indeed, the summation over $m$ is vital in our calculation, as it allows us to apply the addition theorem in Eq.~\eqref{AdditionTheorem} to obtain Eq.~\eqref{Rotalm}. Only after replacing the general $k$-mode with one aligned with the line of sight, $\mathbf k=k\mathbf{e_z}$, the summation over $m$ can be dropped and replaced by the contribution of $m=0$. 

\section{Detailed calculation of $a^{\gamma\pm}_{lm}(\mathbf k)$} \label{Appendixgammakn}

Here, we describe how the expression for $a^{\gamma\pm}_{lm}(\mathbf k)$ given in Eq.~\eqref{almgamma} is calculated. To decompose the shear signal on the sky, we replace the spherical harmonics $Y_{lm}(\mathbf n)$ with the more general spin-weighted spherical harmonics,
\begin{align}
{_s}Y_{lm}=\begin{cases}\sqrt{\frac{(l-s)!}{(l+s)!}}\eth^sY_{lm}\,,\qquad &0\le s\le l\,, \\ \sqrt{\frac{(l+s)!}{(l-s)!}}\bar{\eth}^{-s}Y_{lm}\,,\qquad &0\le -s\le l\,, \\ 0\,,\qquad & l < \vert s\vert\,,\end{cases} \label{defspinweight}
\end{align}
where $\eth$ and $\bar\eth$ are spin-raising and, respectively, -lowering operators defined through
\begin{align}
&\eth{_s}f=-\sin^s\theta\rbr{\partial_\theta+i\sin^{-1}\theta\,\partial_\phi}\sin^{-s}\theta\, f\,,\nnn &\bar\eth{_s}f=-\sin^{-s}\theta\rbr{\partial_\theta+i\sin^{-1}\theta\,\partial_\phi}\sin^{s}\theta\, f\,. \label{defspinop}
\end{align} 
For ease of notation, we also define spin-1 differential operators $\habla_\pm$,
\beeq
\habla_\pm\equiv m^\alpha_\mp\habla_\alpha=\frac{1}{\sqrt{2}}\rbr{\partial_\theta\pm i\sin^{-1}\theta\,\partial_\phi}\,.
\eneq
Assuming that the Fourier mode $\mathbf k$ is aligned with the positive $z$-direction and applying Eq.~\eqref{basisderiv}, we obtain 
\begin{align}
&\mathbf{m}_\pm\cdot \mathbf k=-\frac{k\sqrt{1-\mu^2}}{\sqrt{2}}\,, \qquad \habla_\pm \boldsymbol n=\boldsymbol m_\pm\,, \nnn
&\habla_\pm \boldsymbol m^\pm=\frac{\mu}{\sqrt{2(1-\mu^2)}}\boldsymbol m_\pm\,, \label{C8} 
\end{align}
which we use to calculate
\begin{align}
&\habla_\pm e^{i\bar r\mathbf k\cdot\mathbf n}=i\bar r\mathbf{m_\pm}\cdot\mathbf k \,e^{i\bar r\mathbf k\cdot\mathbf n}=-i\frac{k\bar r\sqrt{1-\mu^2}}{\sqrt{2}}e^{i\bar r\mathbf k\cdot\mathbf n}\,,\nnn
&m^\alpha_\pm m^\beta_\pm \habla_\alpha\habla_\beta
=\habla_\pm^2-\frac{\mu}{\sqrt{2(1-\mu^2)}}\habla_\pm\,,\nnn
&\habla_\pm^2e^{i\bar r\mathbf k\cdot\mathbf n}=-\frac{i\mu k\bar r+k^2\bar r^2(1-\mu^2)}{2}e^{i\bar r\mathbf k\cdot\mathbf n}\,.
\end{align}
From this, it further follows that
\beeq
m^\alpha_\pm m^\beta_\pm\habla_\alpha\habla_\beta e^{i\bar r\mathbf k\cdot\mathbf n}=-\frac{k^2\bar r^2(1-\mu^2)}{2}e^{i\bar r\mathbf k\cdot \mathbf n}\,. \label{usefulrelS}
\eneq
Writing $\Psi(\mathbf n,\bar r)$ as the Fourier transform of $\Psi(\mathbf k,\bar r)$, changing the order of integration and using Eq.~\eqref{usefulrelS}, we obtain
\beeq
{_{\pm 2}}\gamma(\mathbf k,\mathbf n)=-\int_0^{\bar r_s}\mathrm d\bar r\, \frac{(\bar{r}_s-\bar{r})\bar r}{\bar{r}_s}k^2(1-\mu^2)\Psi(\mathbf k,\bar r)e^{i\bar r \mathbf k\cdot \mathbf n}\,, \label{Scalarkn}
\eneq
from Eq.~\eqref{Eq:gammapm} for ${_{\pm 2}}\gamma$.

Now, by applying the definitions of the spin-weighted spherical harmonics and partial integration, we can rewrite the expression for $a_{lm}^{\gamma\pm}(\mathbf k)$ given in Eq.~\eqref{gammadecomp} into the alternative expressions
\begin{align}
&a_{lm}^{\gamma+}(\mathbf k)=\delta_{l\ge 2}\sqrt{\frac{(l-2)!}{(l+2)!}}\int\mathrm d\Omega\, \bar{\eth}^2{_2}\gamma(\mathbf k,\mathbf{ n})Y^\ast_{lm}(\hat{\mathbf n})\,,\nnn
&a_{lm}^{\gamma-}(\mathbf k)=\delta_{l\ge 2}\sqrt{\frac{(l-2)!}{(l+2)!}}\int \mathrm d\Omega \,{\eth}^2{_{-2}}\gamma(\mathbf k,\mathbf{ n})Y^\ast_{lm}(\hat{\mathbf n})\,.\label{apmlm}
\end{align}
Note that the definition of spin-weighted spherical harmonics implies that $a^{\gamma\pm}_{lm}$ is vanishing for $l=0$ and $l=1$, which we express with the symbol $\delta_{l\ge 2}$. 

Applying the spin-lowering and -raising operator twice to ${_{\pm2}}\gamma(\mathbf k,\mathbf n)$, and turning powers of $\mu$ into powers of $-i\partial_x$ acting on $\exp(ix\mu)$, we further obtain
\begin{align}
\bar\eth^2{_{2}}\gamma(\mathbf k,\mathbf n)=\eth^2{_{-2}}\gamma(\mathbf k,\mathbf n) 
=\int_0^{\bar r_s}\mathrm d\bar r \rbr{\frac{\bar{r}_s-\bar{r}}{\bar r_s\bar{r}}}\hat{\mathrm S}(x)\,,
\end{align}
where we have defined the operator
\begin{align}
\hat{\mathrm S}(x)\equiv& 4x^2+x^4+8x^3\partial_x+(12x^2+2x^4)\partial_x^2 \nnn
&+8x^3\partial_x^3+x^4\partial_x^4 \,. \label{Sop}
\end{align}

Finally, using the differential and recursion relations of the spherical Bessel functions to calculate
\beeq
\hat{\mathrm S}(x)j_l(x)=\frac{(l+2)!}{(l-2)!}j_l(x)\,,
\eneq
we can compute $a^{\gamma\pm}_{lm}(\mathbf k)$ from Eq.~\eqref{apmlm} by applying Eq.~\eqref{fundamentaleq}, which yields Eq.~\eqref{almgamma}.


\bibliography{mybib.bib}

\end{document}